\documentclass[preprint,12pt]{elsarticle}

\usepackage[utf8]{inputenc}
\usepackage[T1]{fontenc}
\usepackage{amsmath}
\usepackage{amssymb}
\usepackage{amsthm}
\usepackage{xspace}
\usepackage[colorlinks = true]{hyperref}
\usepackage{lmodern}						
\usepackage{booktabs} 						
\usepackage{siunitx}
\usepackage{csquotes}

\usepackage[shortlabels]{enumitem}

\setlist[description]{align = right,%
					  labelwidth = 2.1cm,%
					  font = \itshape\mdseries,%
					  topsep = .6em,%
					  parsep = 0pt,%
					  itemsep = .5em,%
					  listparindent = 1.2cm}
\setlist[enumerate]{$(i)$,%
					topsep = .5em,%
					parsep = 0pt}				

\usepackage
[
   subrefformat = parens, 
]
{subcaption}

\usepackage
[
    noabbrev,   
    nameinlink, 
]
{cleveref} 

\crefname{cond}{condition}{conditions}
\creflabelformat{stat}{#2{#1}#3}

\crefname{alg}{algorithm}{algorithms}
\creflabelformat{alg}{#2{#1}#3}

\crefname{line}{line}{lines}
\creflabelformat{line}{#2{#1}#3}

\crefname{lem}{lemma}{lemmas}
\creflabelformat{lem}{#2{#1}#3}

\usepackage{dsfont}

\usepackage[algo2e, algoruled, vlined, linesnumbered]{algorithm2e}
\usepackage{setspace}
	
\SetKw{Break}{break}		
\SetKwInput{Input}{Input}
\SetKwInput{Output}{Output}
\SetKw{Proc}{Procedure}
\SetKwFunction{Enum}{enumerate}
\SetKwFunction{Ext}{extendable}

\newtheorem{theorem}{Theorem}
\newtheorem{lemma}[theorem]{Lemma}

\newtheorem{proposition}[theorem]{Proposition}
\newtheorem{conjecture}[theorem]{Conjecture}

\DeclareMathOperator{\littleO}{o}

\DeclareMathOperator{\Or}{O}

\DeclareMathOperator{\rank}{rank}
\DeclareMathOperator{\Tr}{Tr}

\newcommand*{\nwspace}{\hspace*{.1em}}

\newcommand*{\Gyp}{\mathcal{G}}
\newcommand*{\Hyp}{\mathcal{H}}
\newcommand*{\Sys}{\mathcal{S}}
\newcommand*{\Tys}{\mathcal{T}}

\newcommand*{\var}[1]{\texttt{Var}_{#1}}
\newcommand*{\N}{\mathbb{N}}

\newcommand*{\order}{\preccurlyeq}

\newcommand*{\invLex}{\preccurlyeq_{\textup{lex}}}

\newcommand*{\rel}{\mathfrak{r}}

\newcommand{\true}{\textsc{true}\xspace}  
\newcommand{\false}{\textsc{false}\xspace} 
\newcommand*{\FPT}{\textsf{FPT}\xspace}

\renewcommand*{\P}{\textsf{P}\xspace}
\newcommand*{\NP}{\textsf{NP}\xspace}
\newcommand*{\W}{\textsf{W}\xspace}
\newcommand*{\poly}{\textsf{poly}\xspace}

\newcommand*{\ExtHS}{\textup{\textsc{Minimal Hitting Set Extension}}\xspace}
\newcommand*{\extHS}{\textup{\textsc{MinHSExt}}\xspace}
\newcommand*{\MIF}{\textup{\textsc{Multi\-coloured Independent Family}}\xspace}
\newcommand*{\mifabbrv}{\textup{\textsc{MultIndFam}}\xspace}
\newcommand*{\IF}{\textup{\textsc{Independent Family}}\xspace}
\newcommand*{\ifabbrv}{\textup{\textsc{IndFam}}\xspace}
\newcommand*{\WANS}{\textsc{Weighted Antimonotone $3$-normalised Satisfiability}\xspace}
\newcommand*{\wans}{\textsc{WA$3$NS}\xspace}

\newcommand*{\emDash}{\hspace*{.25pt}--\hspace*{.25pt}}

\journal{ArXiv.}

\begin{document}

\begin{frontmatter}

\title{Efficiently Enumerating Hitting Sets of Hypergraphs Arising in Data Profiling\tnoteref{t1}}

\tnotetext[t1]{An extended abstract of this work was presented at the
21st Meeting on Algorithm Engineering and Experiments (ALENEX 2019)~\cite{Blaesius19EfficientlyALENEX}.}

\date{}

\author[1]{Thomas Bl{\"a}sius\fnref{fn1}}
\ead{thomas.blaesius@kit.edu}

\author[2]{Tobias Friedrich\fnref{fn1}}
\ead{tobias.friedrich@hpi.de}

\author{Julius Lischeid\fnref{fn1}}
\ead{jsl71@cantab.ac.uk}

\author[3]{Kitty Meeks\fnref{fn2}}
\ead{kitty.meeks@glasgow.ac.uk}

\author[2]{Martin Schirneck\fnref{fn1}\corref{cor1}}
\ead{martin.schirneck@hpi.de}

\affiliation[1]{organization={Karlsruhe Institute of Technology},
            city={Karlsruhe},
            country={Germany}}

\affiliation[2]{organization={Hasso Plattner Institute, University of Potsdam},
            country={Germany}}

\affiliation[3]{organization={University of Glasgow},
            city={Glasgow},
            country={United Kingdom}}
            
\cortext[cor1]{Corresponding author (\texttt{martin.schirneck@hpi.de}).}

\fntext[fn1]{This work originated while all but the fourth author were
	affiliated with the Hasso Plattner Institute at the University of Potsdam.}
	
\fntext[fn2]{Kitty Meeks is supported by a Personal Research Fellowship
	from the Royal Society of Edinburgh, funded by the Scottish Government.}

\begin{abstract}
	The transversal hypergraph problem is 
	the task of enumerating the minimal hitting sets of a hypergraph.
	It is a long-standing open question whether this can be done in output-polynomial time.
	For hypergraphs whose solutions have bounded size,
	Eiter and Gottlob [SICOMP 1995] gave an algorithm
	that runs in output-polynomial time,
	but whose space requirement also scales with the output size.
	We improve this to polynomial delay and polynomial space.
	More generally, we present an algorithm that on \mbox{$n$-vertex}, $m$-edge hypergraphs 
	has delay $\Or(m^{k^*+1} \nwspace n^2)$ and uses $\Or(mn)$ space,
 	where $k^*$ is the maximum size of any minimal hitting set.
	Our algorithm is oblivious to $k^*$,
	a quantity that is hard to compute or even approximate.
	
	Central to our approach is the extension problem for minimal hitting sets,
	deciding for a set $X$ of vertices whether it is contained in any solution.
	With $|X|$ as parameter,
	we show that this is one of the first natural problems
	to be complete for the complexity class $\W[3]$.
	We give an algorithm for the extension problem running in time $\Or(m^{|X|+1} \nwspace n)$.
	We also prove a conditional lower bound under the Strong Exponential Time Hypothesis,
	showing that this is close to optimal.
	
	We apply our enumeration method to the discovery problem of minimal unique column combinations
	from data profiling.
	Our empirical evaluation suggests that the algorithm outperforms its worst-case guarantees
	on hypergraphs stemming from real-world databases.
\end{abstract}

\begin{keyword}
	data profiling \sep enumeration algorithm
	\sep minimal hitting set \sep transversal hypergraph
	\sep unique column combination \sep W[3]-completeness
\end{keyword}

\end{frontmatter}


\setcounter{footnote}{0}

\section{Introduction}
\label{sec:intro}

\noindent
A recurring computational task in the profiling of relational databases 
is the discovery of hidden dependencies between attributes. 
For instance, \textit{unique column combinations} (UCCs) are subsets of attributes 
such that the value combinations appearing in them are duplicate-free.  
An inclusion-wise minimal UCC reveals structural properties of the stored information
and serves as a small fingerprint of the data.
Unique column combinations, however, are equivalent to \textit{hitting sets} in hypergraphs.
While finding a single minimal hitting set is trivial,
it is usually not enough to decide the existence of a single UCC.
Instead, one aims to compile a comprehensive list of all dependencies.
The UCCs are then used for subsequent data cleaning and enable certain query optimizations~\cite{Abedjan18DataProfiling,Kossmann21Survey}.
One thus has to solve the \textit{transversal hypergraph problem}.
This is the task of enumerating
all minimal hitting sets of a given hypergraph without repetitions.

Besides data profiling, the transversal hypergraph problem also emerges in many other fields,
like artificial intelligence~\cite{Gogic98Recompilation},
machine learning~\cite{DomingoMishraPitt99DualizationLearningMembershipQueries},
distributed systems~\cite{Garcia-Molina85DistributedSystems}, 
integer linear programming~\cite{Boros02MinimalIntegerSolutions},
and monotone logic~\cite{EiterGottlobMakino03NewResults}.
Despite the large interest, the exact complexity of the enumeration problem is still open.
A hypergraph can have exponentially many minimal hitting sets, 
ruling out any polynomial algorithm.
Instead, one could hope for an \emph{output-polynomial} method
whose running time scales polynomially in the input size and the total number of solutions.
Unfortunately, we do not know how to achieve this.
The currently fastest algorithm was presented by Fredman and Khachiyan~\cite{FredmanKhachiyan96Dualization}
and runs in time $N^{\Or(\log N/\log\log N)}$, where $N$ denotes the combined input and output size.
It is the major open question in enumeration, 
whether the transversal hypergraph problem can be solved in output-polynomial time~\cite{DemetrovicsThi87Antikeys,EiterMakinoGottlob08Survey,Mannila87Dependency,Reiter87DiagnosisFirstPrinciples}.

In the absence of a tractable algorithm for general inputs,
special classes of hypergraphs have received a lot of attention.
For example, it is known that
the transversal hypergraph problem admits an output-polynomial algorithm 
when restricted to hypergraphs with bounded edge size~\cite{Boros00EfficientIncremental}
or dual-conformality~\cite{Khachiyan07GlobalParallel},
as well as acyclic hypergraphs~\cite{EiterGottlob95RelatedProblems,EiterGottlobMakino03NewResults}.

In this work, we are interested in the case
where the maximum \emph{solution size} is small.
Given a hypergraph, let $k^*$ be the maximum cardinality of its minimal hitting sets.
This is known as the \emph{transversal rank}.
Indeed, it is very common for hypergraphs arising in data profiling
to have low transversal rank, see~\cite{Kruse16DataAnamnesis,Papenbrock15SevenAlgorithms}.
Eiter and Gottlob~\cite{EiterGottlob95RelatedProblems} 
gave an output-polynomial algorithm for hypergraphs for which $k^*$ is a constant.
We discuss their approach in detail in \Cref{subsec:algorithm_transversal_rank}.
Unfortunately, their algorithm is not usable in data profiling applications 
as its \emph{space} consumption scales with the output size.
Also, one would like to have a guarantee on the \emph{delay}, 
the worst-case time between two consecutive outputs,
that is independent of the number of solutions.
Lastly, although the transversal rank can be expected to be small
usually no a priori bound on $k^*$ is known before the enumeration.
In fact, it is $\W[1]$-hard to compute $k^*$ and \NP-hard to approximate, see~\Cref{subsec:algorithm_transversal_rank}.
We are able to improve in all those aspects 
and obtain an algorithm that is oblivious to $k^*$,
has a space requirement independent of the output size,
and, in the case that $k^*$ is constant, has polynomial delay.

Central to our approach is a subroutine that decides for a set $X$ of vertices
whether it is contained in any minimal hitting set.
We examine the parameterised complexity of this \emph{extension problem},
when parameterised by $|X|$.
We identify it as one of the first natural problems to be complete for the class $\W[3]$.
Prior to the first announcement of this result,
there were only two other problems known with this property.
The first one was given by Chen and Zhang~\cite{ChenZhang06W3W4}
in the context of supply chain management and
Bl{\"a}sius, Friedrich, and Schirneck~\cite{Blaesius16DependencyDetection}
added the detection of inclusion dependencies in relational data.
Since then,
Casel~et~al.~\cite{Casel18ComplexitySolutionExtensionArxiv} 
used the techniques developed in \Cref{sec:extension} to show
$\W[3]$-hardness already for the special case of extension 
to minimal dominating sets in bipartite graphs.
Very recently, Hannula, Song, and Link~\cite{Hannula21IndependenceFromDataArXiv},
building on~\cite{Blaesius16DependencyDetection},
have proven that independence detection in databases is complete for $\W[3]$ as well.

We also approach the extension problem with tools from fine-grained complexity.
Assuming the Strong Exponential Time Hypothesis (SETH),
we prove that our subroutine algorithm is almost optimal.
Moreover, we argue that closing the remaining gap between the upper and lower bound is likely to be
hard, using a nondetermistic extension of SETH recently conjectured by Carmosino et al.~\cite{Carmosino16NSETH}.
Next, we give an overview of our results in detail.

\subsection{Our Contribution}
\label{subsec:intro_contribution}

\noindent
We solve the transversal hypergraph problem with simultaneously 
polynomial delay and space on hypergraphs with bounded transversal rank.
More generally, we devise an algorithm that does not need to know $k^*$.
Notwithstanding, the analysis of the delay depends on the transversal rank.

\begin{theorem}
\label{thm:enum_algorithm}
	There exists an algorithm that on $n$-vertex, $m$-edge hypergraphs
	enumerates the minimal hitting sets with delay $\Or(m^{k^*+1} \nwspace n^2)$ in $\Or(mn)$ space,
	where $k^*$ is the maximum cardinality of any minimal hitting set.\vspace*{.5em}
\end{theorem}

At its core, the algorithm is a tree search in the space of all vertex subsets.
The tree is pruned by deciding for a given set $X$ whether it can be extended to a minimal hitting set.
We analyse the parameterised complexity of this decision with respect to the parameter $|X|$. 

\begin{theorem}
\label{thm:W3-complete}
	The extension problem for minimal hitting sets is complete for $\emph{\W}[3]$
	when parameterised by the cardinality $|X|$ of the set to be extended.
\end{theorem}

It may seem counterintuitive to solve the enumeration problem by reducing
it to a hard decision problem.
The key property we use is that extension is tractable,
provided that $X$ contains only a few vertices.

\begin{theorem}
\label{thm:extension_algorithm}
	There exists an algorithm that decides for an $n$-vertex, $m$-edge hypergraph
	and a set $X$ of vertices
	whether $X$ is contained in any minimal hitting set in time $\Or(m^{|X| + 1} \nwspace n)$
	and space $\Or(mn)$.
\end{theorem}

It is natural to ask whether the exponential dependency on $|X|$ in the running time can be improved.
We give several conditional lower bounds, all of which indicate 
that \Cref{thm:extension_algorithm} is close to optimal.
They present a trade-off between the strength of the conjecture one is willing to assume
and the strength of the resulting bound.

\pagebreak
	
\begin{theorem}
\label{thm:lower_bounds}
	Let $f$ be an arbitrary computable function.
	No algorithm can decide for an $n$-vertex, $m$-edge hypergraph
	and a set $X$ of vertices 
	whether $X$ is contained in any minimal hitting set
	\begin{enumerate}
		\item in time $f(|X|) \cdot \emph{\poly}(m,n)$, unless $\emph{\W}[3] = \emph{\FPT}$;\vspace*{.25em}
		\item in time $f(|X|) \cdot (m\,{+}\,n)^{\littleO(|X|)}$, unless $\emph{\W}[2] = \emph{\FPT}$;\vspace*{.25em}
		\item in time $m^{|X|-\varepsilon} \cdot \emph{\poly}(n)$
			for any constant $|X| \ge 2$ and $\varepsilon > 0$,\vspace*{.25em}
			unless the Strong Exponential Time Hypothesis fails.
	\end{enumerate}
\end{theorem}

\noindent
The SETH-lower bound matches our algorithmic result up to a factor of $m$.
There is a complexity-theoretic obstacle for closing the remaining gap.
We argue that if one could show tight SETH-hardness of the extension problem
with a time bound of $m^{|X|+1-\varepsilon} \cdot \poly(n)$ via a \emph{deterministic} reduction,
this would refute the Nondeterministic Strong Exponential Time Hypothesis (NSETH)~\cite{Carmosino16NSETH}
and thereby resolve several open problems in circuit complexity and satisfiability. 

Finally, we evaluate an implementation of our algorithm 
by applying it to the discovery problem of minimal UCCs.
Our experiments show that our method is much faster on hypergraphs stemming from real-world databases
than the running time bounds would suggest.
In practice, a few simple checks can avoid the worst-case behaviour on many instances,
which boosts the performance.
We also confirm the low memory footprint of our approach.

\subsection{Outline}
\label{subsec:intro_outline}

\noindent
Next, we fix notation and recall basic concepts from combinatorics and complexity theory.
In \Cref{sec:algorithm}, we first review what is known about the transversal rank
and then present our enumeration algorithm.
There, the extension problem for minimal hitting sets is only used as a black box.
It is discussed in detail in \Cref{sec:extension}. 
\Cref{sec:solving_extension} combines the results of the previous two sections
and proves the bounds on the delay and space.
In \Cref{sec:UCC}, we report on the empirical performance of our method in the context of data profiling.
The work is concluded in \Cref{sec:conclusion}.

\section{Preliminaries}
\label{sec:prelims}

\noindent
For a set $S$, let $\mathcal{P}(S)$ be the power set of $S$.
We use $\N^+ = \{1,2, \dots\}$ for the positive integers,
and, for $k \in \N^+$, we set $[k] = \{1, 2, \dots, k\}$.
Computational objects are implicitly assumed to be encoded as bit strings from $\{0,1\}^*$.

\subsection{Hypergraphs and Hitting Sets}
\label{subsec:prelims_HS}

\noindent
A \textit{hypergraph} is a non-empty, finite \textit{vertex set} $V \neq \emptyset$ together
with a system of subsets $\Hyp \,{\subseteq}\, \mathcal{P}(V)$, the \mbox{\textit{(hyper-)edges}}.
A hypergraph is identified with its edge set $\Hyp$ if this does not create ambiguities.
We do not exclude special cases like the empty hypergraph ($\Hyp = \emptyset$),
an empty edge ($\emptyset \in \Hyp$), or isolated vertices ($V \supsetneq \bigcup_{E \in \Hyp} E$).
The  number of edges is $n = |V|$, the number of edges $m = |\Hyp|$.
The \textit{rank} of a hypergraph $\Hyp$ is the
maximum cardinality of its edges.
A \emph{graph} is a hypergraph whose edges all have size exactly $2$.

A \textit{transversal} or \textit{hitting set} of a hypergraph $(V,\Hyp)$ is a set $H \subseteq V$
such that $H$ has a non-empty intersection with every edge $E \in \Hyp$.
A transversal is \textit{(inclusion-wise) minimal}
if it does not properly contain any other transversal.
We extensively use the following observation.

\begin{proposition}[Folklore]
\label{prop:witnesses}
	Let $\Hyp$ be a hypergraph and $H$ a hitting set of $\Hyp$.
	Then, $H$ is minimal if and only if every $x \in H$ has a \emph{private edge}
	$E_x \in \Hyp$ such that $E_x \cap H = \{x\}$.
\end{proposition}

The minimal hitting sets of $\Hyp$ form the \textit{transversal hypergraph} $\Tr(\Hyp)$
on the same vertex set $V\!$.
We occasionally denote the number of its edges by $N_{\min}$.
We abbreviate the rank of the transversal hypergraph $\rank(\Tr(\Hyp))$
as \emph{transversal rank}, and,
if $\Hyp$ is clear from the context, we use $k^*$ to denote it.

A hypergraph is \textit{Sperner} if none of its edges is properly contained
in another. The \textit{minimisation} of $\Hyp$ is the subhypergraph of
inclusion-wise minimal edges, 
$\min(\Hyp) = \{ E \in \Hyp \mid \forall E' \in \Hyp \colon E' \subseteq E \Rightarrow E' = E\}$.
Note that $\min(\Hyp)$ and $\Tr(\Hyp)$ are always Sperner hypergraphs.
Regarding transversals, it does not make a difference whether
the full hypergraph is considered or its minimisation as we have $\Tr(\Hyp) = \Tr(\min(\Hyp))$.
Moreover, the minimisation and transversal hypergraph are mutually dual,
meaning $\Tr(\Tr(\Hyp)) = \min(\Hyp)$.
For any two Sperner hypergraphs $\Gyp$ and $\Hyp$,
we have $\Gyp = \Tr(\Hyp)$ if and only if $\Hyp = \Tr(\Gyp)$.
The minimisation is computable in quadratic time.
In this work, we therefore assume all hypergraphs to be Sperner.

Any total ordering $\order$ of the vertex set $V$ induces a \textit{lexicographical order}
on $\mathcal{P}(V)$.
Following the definition in~\cite{Johnson88MaxIndSet},
we say a subset $S \subseteq V$ is \emph{lexicographically smaller (or equal)} than subset $T$,
denoted $S \invLex T$,
if either they are equal or the \mbox{$\order$-smallest} element in which $S$ and $T$ differ is in $S$.
We call a hypergraph on a totally ordered vertex set an \textit{ordered hypergraph}

\subsection{Parameterised Complexity and the Strong Exponential Time Hypothesis}
\label{subsec:prelims_FPT}

\noindent
The \emph{decision problem} of a set $\Pi \subseteq \{0,1\}^*$
is  to answer for an \emph{instance} \mbox{$I \in \{0,1\}^*$} whether $I \in \Pi$.
Such a problem is \emph{parameterised} if $I$
comes augmented with a non-negative integer \emph{parameter} $k$.
We then have $\Pi \subseteq \{0,1\}^* \times \N^+$.
A parameterised  problem is \textit{fixed-parameter tractable} (FPT),
if there exists a computable function $f \colon \N^+ \to \N^+$ such that the input $(I,k)$
can be decided in time $f(k) \,{\cdot}\, \poly(|I|)$.
The class of all fixed-parameter tractable problems is denoted by \FPT.
Slightly abusing notation, we say any algorithm (not only for decision problems)
that takes time $f(k) \,{\cdot}\, \poly(|I|)$ runs in  \textit{FPT-time}.
We mostly employ more expressive quantizations of the input size,
which are still polynomially related to the encoding length $|I|$.
For example, if the instance $I = (V,\Hyp)$ is a hypergraph,
we use $n = |V|$ and $m = |\Hyp|$.

Let $\Pi$ and $\Pi'$ be parameterised problems.
A \emph{parameterised reduction} from $\Pi$ to $\Pi'$ is a function
computable in FPT-time that maps an instance $(I, k)$ of $\Pi$ to an
equivalent instance $(I', k')$ of $\Pi'$
such that there is a computable function $g$ with $k' \le g(k)$.
A parameterised reduction is called \emph{linear}~\cite{Chen06LowerBounds}
if $g$ is a linear function, that is, if $k' = \Or(k)$.
All parameterised reductions we give in this work are linear.

Parameterised reductions give rise to a hierarchy of complexity classes, the \emph{\emph{\W}-hierarchy}.
There are several equivalent ways to define it~\cite{FlumGrohe06ParameterizedComplexityTheory};
we choose the one in terms of circuits.
A \emph{(Boolean) circuit} is a directed acyclic graph whose vertex set consists of input
nodes, NOT-, AND-, and OR-gates, with the obvious semantics, and a single output node.
AND- and OR-gates have potentially unbounded fan-in.
A \emph{(Boolean) formula} is a circuit in which every gate has fan-out $1$.
The \textit{depth} of a circuit is the maximum length of a path from an input to the output node. 
The \textit{weft} is the maximum number of \emph{large gates} with fan-in larger than $2$ on any path.
The \textsc{Weighted Circuit Satisfiability} problem is to decide
for a given circuit $C$ and a positive integer $k$
whether $C$ has a satisfying assignment of \textit{(Hamming) weight $k$},
that is, with exactly $k$ input nodes set to \true.
The parameter is $k$.
The class $\W[\P]$ is the collection of all parameterised problems
that admit a parameterised reduction to \textsc{Weighted Circuit Satisfiability}.
Analogously, for or any positive integer~$t$,
the \textsc{Weighted Circuit Satisfiability} restricted to 
circuits of constant depth and weft at most $t$ is the defining complete problem for the class $\W[t]$.
The classes
$\FPT \subseteq \W[1] \subseteq \W[2] \subseteq \W[3] \subseteq \dots \subseteq \W[\P]$
form the $\W$\emph{-hierarchy}.
All inclusions are conjectured to be strict, 
see~\cite{DowneyFellows13Parameterized,Niedermeier06Invitation}

Another source of conditional lower bounds is the 
\textit{Strong Exponential Time Hypothesis} (SETH)~\cite{Impagliazzo01ETH2}.
It states that, for every $\varepsilon > 0$, there exists a positive integer $k = k(\varepsilon)$
such that no algorithm can decide the satisfiability of
Boolean formulas in conjunctive normal form with $k$ literals per clause ($k$-CNF~SAT) on $n$ variables 
in time $\Or(2^{(1-\varepsilon)n})$.
A weaker assumption is
the \emph{Exponential Time Hypothesis} (ETH)~\cite{Impagliazzo01ETH1,Impagliazzo01ETH2}
that $3$-CNF~SAT cannot be solved in time $2^{\littleO(n)}$.
ETH implies that the $\W[1]$-complete \textsc{Independent Set} problem
on $n$-vertex, $m$-edge graphs cannot be solved
in time \mbox{$f(k) \cdot (m+n)^{\littleO(k)}$},
whence $\W[t] \neq \FPT$ for all $t \ge 1$~\cite{Chen06LowerBounds}.

\subsection{Enumeration Complexity}
\label{subsec:prelims_enumeration}

\noindent
It is enough for our purposes to define \emph{enumeration} informally as the task
of computing and outputting all solutions to a computational problem without repetition.\footnote{
	Enumeration should not be confused with merely counting the number of solutions.
}
We are only concerned with the \emph{transversal hypergraph problem},
that is, given a hypergraph $\Hyp$, enumerating the edges of $\Tr(\Hyp)$.
An \textit{output-polynomial} enumeration algorithm runs in time polynomial in both the input and output size.
That means the enumeration succeeds within $\poly(n,m,N_{\min})$ steps.
A seemingly stronger requirement is an \textit{incremental polynomial} algorithm,
generating the solutions in such a way that the $i$-th
\textit{delay}, the time between the $(i{-}1)$-st and $i$-th output, is in $\poly(n,m,i)$.
This includes the preprocessing time until the first solution arrives ($i=1$)
and the postprocessing time between the last solution and termination ($i=N_{\min} + 1$).
It is known that the transversal hypergraph problem can be solved in output-polynomial time
if and only if it admits an incremental polynomial algorithm~\cite{Bioch95Identification}.
This is not necessarily true for other enumeration problems.
The strongest form of output-efficiency is that of \textit{polynomial delay}, where the
delay is universally bounded by a polynomial in the input size only.
One can also restrict the space consumption.
Ideally, the algorithm only uses space polynomial in the input.
Even if $N_{\min}$ is guaranteed to be polynomial in $m$ and $n$,
the space should be independent of $N_{\min}$.

\subsection{Relational Databases and Unique Column Combinations}
\label{subsec:prelims_RelData}

\noindent
To describe relational data, we fix a non-empty, finite \textit{(relational) schema} $R$.
The elements of $R$ are the \textit{attributes} or \textit{columns} and
each attribute comes implicitly associated with a set of admissible values.
\textit{Rows} over $R$ are tuples $r$ whose entries are indexed
by $R$ such that, for each $a \in R$, the \emph{value} $r[a]$ is admissible for attribute $a$.
For a set $X \subseteq R$ of columns, we let $r[X]$ denote the subtuple of
$r$ consisting only of the entries indexed by $X$.
A  \textit{(relational) database} $\rel$ over $R$ is a finite set of rows.

In some database $\rel$ over schema $R$,
a set $X \subseteq R$ is a \emph{unique column combination} (UCC) if for any two
distinct rows $r,s \in \rel$, $r \neq s$, we have $r[X] \neq s[X]$.
A UCC is \textit{(inclusion-wise) minimal} if it does not properly contain any other UCC. 
There is a one-to-one correspondence between UCCs and transversals.
Let $r,s \in \rel$ be distinct rows and $\{a \in R \mid r[a] \neq s[a] \}$
their \emph{difference set}.
Then, the (minimal) UCCs are exactly the (minimal) hitting sets
of the hypergraph of difference sets for all pairs of rows in $\rel$.

\section{Enumerating Minimal Hitting Sets}
\label{sec:algorithm}

\noindent
In this section, we outline our enumeration algorithm for minimal hitting sets.
Our main motivation comes from data profiling,
so we design our method with an eye on instances that have small solutions.
Nevertheless, we aim for a general-purpose algorithm 
and do not restrict the possible input hypergraphs.
Therefore, the algorithm does not make any assumptions on the inputs
and relies only on the given hypergraph itself.
Its analysis, however, incorporates the transversal rank.
%
%
Before we present our algorithm, we discuss some alternative approaches
and asses how we can improve upon them. 

\subsection{On the Transversal Rank}
\label{subsec:algorithm_transversal_rank}

\noindent
We review here what is known algorithmically about the transversal rank $k^*$.
This also serves to highlight the subtle differences
in measuring the complexity of an enumeration algorithm.
%
Eiter and Gottlob~\cite{EiterGottlob95RelatedProblems} showed that
the transversal hypergraph problem can be solved in incremental polynomial time
on instances for which $k^*$ is bounded.
Their result hinges on the following proposition.

\begin{proposition}[Eiter and Gottlob~\cite{EiterGottlob95RelatedProblems}]
\label{prop:inc_poly}
	Let $\Hyp$ and $\Gyp$ be two hypergraphs on the same vertex set $V\!$ 
	and let $k = \rank(\Gyp)$.
	There exists an algorithm that decides whether $\Gyp = \Tr(\Hyp)$ in time
	$\Or( |\Hyp| |\Gyp| |V| + (|\Hyp| {+} |\Gyp|)|V|^{k+1}  + |\Hyp|^{k+2} |V|)$
	and space $\Or((|\Hyp| {+} |\Gyp|)|V|)$.
	Moreover, if $\Gyp \subsetneq \Tr(\Hyp)$, the algorithm finds
	a minimal transversal $T \in \Tr(\Hyp){\setminus}\Gyp$ within the same bounds.
\end{proposition}

The enumeration starts with the empty hypergraph $\Gyp = \emptyset$
and repeatedly checks whether $\Gyp = \Tr(\Hyp)$, that is, 
whether $\Gyp$ already contains all solutions.
If not, a new solution $T \notin \Gyp$ is computed.
Note that $\Gyp \subseteq \Tr(\Hyp)$ is an invariant,
whence $k = \rank(\Gyp)$ is always at most $k^* = \rank(\Tr(\Hyp))$.
However, this approach has two drawbacks.
It is already unfortunate that the delay depends on $|\Gyp|$ and thus on $|\!\Tr(\Hyp)|$,
but it is indeed prohibitive in practice 
that the space consumption scales with the number of solutions.

If one is working with a class of hypergraphs for which one suspects $k^*$ to be small,
albeit no a priori bound is known, one could be tempted to compute the transversal rank first 
and then brute-force all sets up to that size.
Computing $k^*$ is \NP-hard~\cite{Cheston90UpperFractionalDomination,ColombNourine08KeysofFormalContext}.
Bazgan~et~al.~\cite{Bazgan18FacetsofUpperDomination} further showed that 
it is $\W[1]$-hard, parameterised by $k^*$,
and that $k^*$ cannot be approximated within a factor of $n^{1-\varepsilon}$
for any constant $\varepsilon>0$, unless $\P = \NP\!$.

The parameterised hardness stems from the potentially unbounded size of the hyperedges.
Fernau~\cite{Fernau05ParameterizedAlgorithmics} showed that the transversal rank of a graph\footnote{%
	This is more commonly known as \textsc{Maximum Minimal Vertex Cover}~\cite{Boria15MaxMinVertexCover,Fernau05ParameterizedAlgorithmics}.
} can be computed in FPT-time by presenting an algorithm running in time $\Or(2^{k^*}) + \poly(n)$,
which was later improved to $1.5397^{k^*} {\cdot}\, \poly(n)$~\cite{Boria15MaxMinVertexCover}.
For an arbitrary constant $c$,
Damaschke~\cite{Damaschke11DoubleHypergraphDualization} used \Cref{prop:inc_poly} to give an algorithm
that computes the transversal rank of a hypergraph whose edges have at most $c$ vertices 
in time $c^{k^*} {\cdot}\, p_c(m,n)$,
where $p_c$ is a polynomial whose degree depends on $c$.
If $c$ is seen as another parameter
(namely, computing $k^*$ parameterized by $c+k^*$),
there exists an FPT-algorithm running 
in time $2^{ck^*} {\cdot}\, \poly(m,n)$~\cite{Araujo21MaximumMinimalBlockingSet}.

Returning to hypergraphs with unbounded edge size, the parameterized reduction in~\cite[Theorem~23]{Bazgan18FacetsofUpperDomination} that shows the $\W[1]$-hardness of computing the transversal rank
has a quadratic blowup in the parameter.
The lower bound on \textsc{Independent Set} by Chen et al.~\cite{Chen06LowerBounds} 
(see \Cref{subsec:prelims_FPT})
thus implies that $k^*$ cannot be computed in time \mbox{$f(k^*) (m+n)^{\littleO(\sqrt{k^*})}$}
for any computable function $f$, unless the Exponential Time Hypothesis fails.
Very recently and independently of each other, Araújo et al.~\cite{Araujo21MaximumMinimalBlockingSet}
as well as Dublois, Lampis, and Paschos~\cite{Dublois21UpperDominatingSet}
raised the bound to
\mbox{$f(k^*) (m+n)^{\littleO({k^*})}$}.
This essentially matches the currently fastest algorithm,
which uses the following characterization of the transversal rank by Berge and Duchet~\cite{BergeDuchet75GilmoresTheorem}.
For a hypergraph $(V,\Hyp)$, subhypergraph $\Hyp' \subseteq \Hyp$, and vertex $v \in V$,
let $\deg_{\Hyp'}(v) = | \{E \in \Hyp' \mid v \in E\} |$ denote the \emph{degree of $v$ in $\Hyp'$}.

\begin{proposition}[Berge and Duchet~\cite{Berge89Hypergraphs,BergeDuchet75GilmoresTheorem}]
\label{prop:Berge_Duchet}
	Let $(V,\Hyp)$ be a Sperner hypergraph and $k \ge 2$ an integer.
	The transversal rank of $\Hyp$ is at most $k$ if and only if,
	for all subhypergraphs $\Hyp' \subseteq \Hyp$ with $|\Hyp'| = k+1$ edges,
	there exists an edge of $\Hyp$ that is contained in the set $\{v \in V \mid \deg_{\Hyp'}(v) > 1\}$.
\end{proposition}

Recall from \Cref{subsec:prelims_HS} that the assumption
of the input hypergraph being Sperner does not loose generality.
One can thus test the condition of \Cref{prop:Berge_Duchet} for increasing $k$.
The value that satisfies it for the first time is $k^*$.
The last iteration dominates the running time, which gives a bound of
$\Or\!\big(\binom{m}{k^*+1} (k^*\,{+}\, m) \nwspace n \big)
	= \Or(m^{k^*+2} \nwspace n)$.
The subsequent test of the sets with up to $k^*$ vertices adds another $\Or(m n^{k^*})$ term.
The enumeration time is polynomial for bounded $k^*$,
but the algorithm does not admit any non-trivial guarantees on the delay.
Also, the space requirement again depends on the total number of solutions
since the algorithm has to avoid testing supersets of minimal solutions.

In contrast, we give an algorithm with delay $\Or(m^{k^*+1} \nwspace n^2)$, 
which is better than the bound above for $m > n$.
More importantly though, our algorithm uses space that is only linear in the input size regardless of $k^*$.

\subsection{Backtracking Enumeration with an Extension Oracle}
\label{subsec:algorithm_backtracking}

\noindent
It is a common pattern in the design of enumeration algorithms to base
them on a so-called \textit{extension oracle} as introduced by Lawler~\cite{Lawler72KBestSolutions}.
The oracle, tailored to the combinatorial problem at hand,
is queried with a set of elements of the underlying universe 
and decides whether there exists a solution that contains these elements.
Applications of this technique usually involve
settings in which the extension problem is solvable in polynomial time,
like for cycles and spanning trees~\cite{Read75BacktrackAlgorithm}, 
motif search in graphs~\cite{Bjoerklung15MotifSearch},
or satisfying assignments for restricted fragments of propositional logic~\cite{Creignou97GeneralizedSATProblems}.
For us, the situation is different
in that the extension problem for minimal hitting sets is \NP-complete~\cite{Boros98Subimplicants}.
We show later in \Cref{sec:extension} that the problem is also hard in a parameterised sense.
At first, it may seem paradoxical that reducing enumeration to a hard decision problem
can speed up the resulting algorithm.
We exploit the fact that the time needed to solve the extension problem 
is small (enough) for sets that contain only a few vertices.

The original oracle technique~\cite{Lawler72KBestSolutions} 
consists of fixing certain elements of the partial solution
and then extending it to the optimum, with respect to a certain ranking function,
among all objects that share the fixed elements.
During the computation the new candidates are stored in a priority queue.
The main bottleneck is the space demand of the queue.
For every partial solution, the number of newly introduced candidates
can be equal to the size of the universe, meaning exponential growth.
Therefore, modifications are necessary to implement the technique efficiently.

In addition to the extension oracle,
we use a decision tree to guide the search for minimal solutions
in the power lattice of all subsets of the universe.
This is known as \textit{backtracking enumeration}~\cite{Read75BacktrackAlgorithm}
or \textit{flashlight technique}~\cite{Mary16Closure}.
It allows us to reduce the space requirement to only polynomial in the input.

In the following, we show how to combine both ideas.
Let $(V,\Hyp)$ be a hypergraph.
Suppose we are given an oracle that, queried with disjoint sets $X,Y \subseteq V\!$,
answers whether there exists a minimal solution $T \in \Tr(\Hyp)$
such that $X \subseteq T \subseteq V{\setminus}Y$,
that is, whether $X$ is extendable avoiding $Y$.
We use this to enumerate all such $T$.
If $X \cup Y = V\!$, this can only be $T = X$ itself.
Otherwise,
we recursively compute the solutions for the pairs $(X \,{\cup}\, \{v\},Y)$ and $(X,Y \,{\cup}\, \{v\})$,
where $v$ is a vertex neither contained in $X$ nor $Y$.
In other words, we (implicitly) build a binary tree
whose nodes are labelled with the pairs $(X,Y)$.
The node $(\emptyset,\emptyset)$ is the root and
the children of $(X,Y)$ are $(X \,{\cup}\, \{v\},Y)$ and $(X,Y \,{\cup}\, \{v\})$.
Let $\order$ be a total order on $V\!$.
Always choosing the $v$ as the $\order$-smallest element of \mbox{$V{\setminus}(X \cup Y)$}
gives a universal branching order.
This obviates the need of additional communication between the nodes or any shared memory.
It is another ingredient to reduce the space demand.
In particular, we do not need to record previously found solutions to guide the search.
Distinct branches of the tree are independent making the algorithm trivially parallelisable.
This is, however, not the focus of this work.

In the absence of any pruning, the recursion would produce the full binary tree 
with leaves $(X,V{\setminus}X)$ for \textit{every} possible set $X \in \mathcal{P}(V)$.
However, we only need to enter the subtree if one of its leaves is labelled with a minimal hitting set.
For the subtree rooted in $(X,Y)$,
this is the case iff $X$ can be extended to a minimal hitting set without the vertices in $Y$.

\begin{algorithm2e}
\setstretch{1.1}
\vspace*{.25em}
	\KwData{Non-empty ordered hypergraph $(V,\order,\Hyp)$.}
	\Input{Partition $(X, Y, R)$ of the vertex set $V$.}
	\Output{The minimal hitting sets $T \in \Tr(\Hyp)$ with $X \subseteq H \subseteq V{\setminus}Y$.}
	\vspace*{.5em}
	\Proc{\Enum{$X, Y, R$}}{:\\
		\vspace*{.5em}
		\lIf{$R = \emptyset$} {\Return $X$}
		$v \gets \min_{\order} R$\;
		\emph{isExtendable} $\gets\ $\Ext{$X {\cup} \{v\}, Y$}\; \vspace*{.25em}
		\lIf{ isExtendable $==$ \emph{\textsc{minimal}}}{\Return $X \cup \{v\}$} \label{line:zeroth_Ext}
		\lIf{ isExtendable $==$ \emph{\textsc{true}} \label{line:first_Ext}}
			{\Enum{$X \cup \{v\}, Y, R {\setminus} \{v\}$}}	
		\lIf{ \emph{(}isExtendable $==$ \emph{\textsc{false}}
			\emph{\textbf{or}} \Ext{$X, Y \cup \{v\}$} $==$ \emph{\textsc{true}}\emph{)} \label{line:second_Ext}}
		{\Enum{$X, Y \cup \{v\}, R {\setminus} \{v\}$}}		
	}	
\caption{Recursive algorithm for the Transversal Hypergraph problem.
	The initial call is \mbox{\texttt{enumerate(}$\emptyset$, $\emptyset$, $V$\texttt{)}}.}
	\label[alg]{alg:enumerate_recursively}
\end{algorithm2e}

We formalise this approach in \Cref{alg:enumerate_recursively}.
Assume for now that subroutine \Ext{$X, Y$} solves the extension problem
for the given pair of sets
and additionally reports if $X$ itself is already a minimal hitting set.
Namely, it returns \textsc{minimal} if $X \in \Tr(\Hyp)$,
\true if there exists some $T \in \Tr(\Hyp)$ with $X \subsetneq T \subseteq V{\setminus}Y$,
and \false otherwise.
We defer the implementation details of \Ext to \Cref{sec:solving_extension}.
Procedure \Enum handles the work inside a node of the decision tree.
Besides the depth-first, pre-order traversal of the tree,
it also exercises two short-cut evaluations.
For this, the ternary variable \emph{isExpendable} holds the result of the first check.
If the set $X \cup \{v\}$ is a minimal solution, we output it and return.
If it cannot be extended,
we immediately recurse on the right-child \emph{without} calling the
potentially expensive second check \Ext{$X, Y \cup \{v\}$}.
The initial call is \mbox{\Enum{$\emptyset$, $\emptyset$, $V$}}.

\begin{lemma}
\label{lem:enumerate_lexicographically}
	Let $(V,\Hyp,\order)$ be a non-empty ordered hypergraph.
	Suppose, for disjoint sets $X,Y \subseteq V$, subroutine \Ext{$X, Y$} decides
	whether there exists a $T\in \Tr(\Hyp)$ with $X \subseteq T \subseteq V{\setminus}Y$ and,
	if so, whether $X = T$.
	Then, \Cref{alg:enumerate_recursively} enumerates the edges of $\Tr(\Hyp)$
	in $\order$-lexicographical order.
\end{lemma}

\begin{proof}
	The correctness is almost immediate from the discussion above.
	Only the shortcut evaluations have not yet been argued.
	If the set $X \cup \{v\}$ is not only extendable without $Y\!$, but even minimal itself,
	then adding any more vertices from $V{\setminus}(X \cup Y \cup \{v\})$
	will make it unextendable.
	Adding these vertices to $Y$ instead does not change $X \cup \{v\}$ being a minimal solution.
	In summary, we already know in advance the outcomes of all extension checks in the whole subtree
	rooted in $(X \cup \{v\},Y)$.
	The set $X \cup \{v\}$ is the only solution that remains in that tree and 
	we can safely output it and backtrack.
	
	Regarding the second shortcut in line~\ref{line:second_Ext},
	the recursion enters the node $(X,Y)$ only if there exists a
	$T\in \Tr(\Hyp)$ with $X \subsetneq T \subseteq V{\setminus}Y$.
	If the first check \Ext{$X \cup \{v\}$, $Y$} returns \false,
	no such $T$ contains the vertex $v$.
	Instead, all solutions occur in the subtree rooted at the right child 
	$(X, Y \cup \{v\})$ and we do not need to perform the second evaluation.
	Note that \Ext{$X$, $Y \cup \{v\}$} cannot return \textsc{minimal}
	due to $X \neq T$.
	
	In the extreme case of $\Hyp$ having not a single hitting set
	(that is, $\emptyset \in \Hyp$)
	both checks in lines~\ref{line:first_Ext} and \ref{line:second_Ext} 
	fail already in the root node.
	The algorithm then returns immediately without an output.
	Here, we use the assumption that $\Hyp$ has at least one vertex
	and thus $R = V \neq \emptyset$ holds in the root.
	
	Finally, we prove that the algorithm outputs the minimal transversals in lexicographical order.
	First, observe that the labelling of the nodes is injective
	as it encodes the unique path from the root. 
	To see this, let $v_1 \order v_2 \order \dots \order v_n$ be the total order.
	Any node with distance $k$ to the root has $X \cup Y = \{v_1, \dots, v_k\}$
	and $X$ contains exactly those branching nodes at which the recursion entered the left child.
	Now let $a = (X_a,Y_a)$ and $b = (X_b,Y_b)$ be two distinct leaves
	such that the pre-order traversal visits $a$ before $b$.
	We have $X_a \neq X_b$ from the injective labelling, whence the symmetric difference
	$X_a \,\triangle\, X_b = (X_a{\setminus}X_b) \cup (X_b{\setminus}X_a)$ is non-empty.
	Define $v = \min_{\order} X_a \,\triangle\, X_b$.
	This is the branching vertex of the lowest common ancestor of $a$ and $b$.
	\Cref{alg:enumerate_recursively} first tries to add $v$ to the current partial solution
	 in line~\ref{line:first_Ext}, from which $v \in X_a$ and $X_a \invLex X_b$ follow.
\end{proof}

\Cref{alg:enumerate_recursively} bears some similarity to the backtracking method
by Elbassioni, Hagen, and Rauf~\cite[Figure~1]{ElbassioniHagenRauf08FPTHypergraphDuality}.
The main difference is the search for new solutions.
In our algorithm, the nodes in the decision tree
maintain the partial solution $X$ and additionally the set $Y$ of vertices
that have already been excluded.
The branching vertex $v$ is chosen, somewhat arbitrarily, by the order $\order$.
In contrast, the algorithm in~\cite{ElbassioniHagenRauf08FPTHypergraphDuality}
works only on the partial solution $X$ and
explicitly computes a new vertex to extend it, which is computationally expensive.
Also, their check whether $X$ is already minimal is redundant.
This information can be obtained as a by-product of a careful
implementation of \Ext at no extra cost, see \Cref{lem:algorithm_ExtHS}.

We employ the order on the vertex set to reduce the need for coordination during the search.
The induced lexicographic order on the outputs can also be useful in the application domain.
For example in the context of data profiling, 
it ensures that ``interesting'' unique column combinations are discovered first.
Suppose the attributes of a database are ranked by importance, 
then the lexicographic enumeration starts with those combinations that contain many important attributes.
However, the order also raises some complexity-theoretic issues. 
Computing the lexicographically smallest minimal hitting set is
\mbox{\NP-hard}~\cite{Eiter94ExactTransversal,Johnson88MaxIndSet}.
Therefore, it is unlikely that any implementation of the extension subroutine can lead to \Cref{alg:enumerate_recursively} having polynomial
delay on \emph{all} ordered hypergraphs.
Notwithstanding, we present an implementation 
such that our algorithm achieves polynomial delay at least on instances with bounded transversal rank.
We also evaluate the impact of the order
on the empirical run time on real-world databases in our experiments in \Cref{sec:UCC}.

\section{Minimal Hitting Set Extension}
\label{sec:extension}

\noindent
We previously assumed an oracle deciding 
whether a set of vertices can be extended to a minimal hitting sets.
Here, we examine the computational hardness of this decision.
The insights gained here will later lead to an algorithm for the subroutine
with an almost optimal running time.

It is easy to see that,
for a hypergraph $(V,\Hyp)$ and disjoint sets $X,Y \subseteq V$,
there exists a minimal transversal $T \in \Tr(\Hyp)$ such that $X \subseteq T \subseteq V{\setminus}Y$
if and only if the \emph{truncated} hypergraph 
$\Hyp' = \{E{\setminus}Y \mid E \in \Hyp\}$ has a minimal hitting set $T$ with $X \subseteq T$.
Indeed, the witnessing transversal $T$ is the same for both $\Hyp$ and $\Hyp'$.
We thus define the extension problem as follows.

\vspace*{.5em}
\noindent
\ExtHS (\extHS)
\begin{description}
	\item [Instance:] A hypergraph $(V,\Hyp)$ and a sets $X \subseteq V\!$.
	\item [Parameter:] The cardinality $|X|$.
	\item [Decision:] Is there a minimal hitting set $T$ of $\Hyp$ with $X \subseteq T$?
\end{description}

Boros, Gurvich, and Hammer showed that the unparameterised variant of
\extHS is \NP-complete in general but tractable if $|X|$ is bounded~\cite{Boros98Subimplicants}.
This and the fact that minimal hitting sets in many applications are small
warrants a parameterised investigation with respect to the cardinality $|X|$.
Observe that \extHS generalises the extension problem for minimal vertex covers in graphs.
Casel et al.~\cite{Casel19Extension} proved $\W[1]$-completeness of the latter.

In~\cite{Boros98Subimplicants},
\extHS was reduced to a certain covering problem in hypergraphs.
We extend this result by proving that the extension and covering problems are in fact
equivalent under parameterised reductions.
We then use this equivalence to show that \ExtHS is one of the first natural problems
to be complete for the parameterised complexity class $\W[3]$.
We further prove conditional lower bounds
on the running time of any algorithm for the extension problem,
assuming that certain collapses in the $\W$-hierarchy do not occur
or that the Strong Exponential Time Hypothesis is true, respectively.

\subsection{$\emph{\W}[3]$-Completeness}
\label{subsec:extension_W3}

\noindent
We present necessary and sufficient conditions for a set of vertices
to be a subset of some minimal hitting set.
This naturally extends the characterisation of minimal transversals in \Cref{prop:witnesses}.
The result appears implicitly in~\cite{Boros98Subimplicants},
we give a self-contained proof below.

\begin{proposition}[Boros, Gurvich, and Hammer~\cite{Boros98Subimplicants}]
\label{prop:char_extension}
	Let $(V,\Hyp)$ be a hypergraph and $X \subseteq V$ a set of vertices.
	There is a $T \in \Tr(\Hyp)$ with $X \subseteq T$ if and only if
	there exists a family of edges $\{E_x\}_{x \in X} \subseteq \Hyp$ such that
	\begin{enumerate}
		\item\label[cond]{cond:char_ext_private} 
			for every vertex $x \in X$, we have $E_x \cap X = \{x\}$;
		\item\label[cond]{cond:char_ext_covering} 
			for every edge $E \in \Hyp$ contained in $\bigcup_{x \in X} E_x$,
			we have $E \cap X \neq \emptyset$.
	\end{enumerate}
\end{proposition}

\begin{proof}
	Let $T$ be a minimal hitting set that contains $X$.
	\Cref{prop:witnesses} guarantees a private edge $E_x \in \Hyp$ with respect to $T$
	for every $x \in X$.
	Let further $E \in \Hyp$ be such that $E \subseteq \bigcup_{x \in X} E_x$.
	As $T$ is a hitting set, there exists a vertex $y \in E \cap T$.
	From $(\bigcup_{x \in X} E_x) \cap T = X$, we conclude that $y$ must be in $X\!$.
	Hence, the private edges also fulfil \Cref{cond:char_ext_covering}.
	
	Conversely, suppose $\{E_x\}_{x \in X}$ is a suitable collection of hyperedges.
	\Cref{cond:char_ext_covering} implies that $H = X \cup (V{\setminus}\bigcup_{x \in X} E_x)$ 
	is a (not necessarily minimal) hitting set of $\Hyp$.
	Let $T \subseteq H$ be any \emph{minimal} hitting set,
	then $T$ contains every $x \in X$ as otherwise $E_x$ would not intersect $T$ by
	\Cref{cond:char_ext_private}.
\end{proof}

We call an edge $E$ a \emph{candidate private edge} for $x \in X$ (with respect to set $X$)
if $E \cap X = \{x\}$ holds.
The partial solution $X$ has some extension $T \in \Tr(\Hyp)$ iff 
there is a collection of candidate private edges $\{E_x\}_{x \in X}$
that satisfy \Cref{cond:char_ext_covering}.
Then, the $E_x$ indeed serve as private edges with respect to $T$ 
in the sense of \Cref{prop:witnesses}.

In light of this characterisation, we define an intermediate parameterised problem,
which we call \MIF.
It captures the following computational task:
given $k$ lists of sets together with an additional collection of ``forbidden'' sets,
one has to select one set from each list such that they do not completely cover any forbidden set.

\vspace*{.5em}
\noindent
\MIF (\mifabbrv)
\begin{description}
	\item [Instance:] A $(k{+}1)$-tuple $(\Sys_1,\ldots, \Sys_k, \Tys)$ of hypergraphs
	
	$\,$on the common vertex set $U$.
	\item [Parameter:] The non-negative integer $k$.
	\item [Decision:] Are there edges $S_1 \in \Sys_1, \ldots, S_k \in \Sys_k$ 
		such that $\bigcup_{i = 1}^k S_i$
	
		\hspace*{.9mm}does not contain an edge of $\Tys$?
\end{description}
\vspace*{.25em}

The \MIF problem generalises \textsc{Multicoloured Independent Set} on graphs
where the vertex set is partitioned into $k$ ``colour classes'' and
the desired independent set is required to contain one vertex of each colour~\cite{Cygan15ParamertizedAlgorithms,Fellows09MultiInterval}.
In the generalisation, we instead select \textit{sets} of vertices
such that their \textit{union} has to be independent.
Now the sets have ``colours'' and the  $\Sys_i$ represent the colour classes.
\textsc{Multicoloured Independent Set} is the special case
in which the hypergraphs $\Sys_i$ consist entirely of singletons and $\Tys$ of the edges of the graph.
Evidently, \MIF is $\W[1]$-hard.

We now prove the equivalence between \mifabbrv and \extHS.
We report the features of the second reduction in full detail
as we need them later for the fine-grained lower bounds.

\begin{lemma}
\label{lem:extension_and_multIndFam}
	\ExtHS and \MIF are equivalent under linear parameterised reductions.\vspace*{.5em}
	
	\noindent
	The reduction to the \ExtHS problem
	takes time $\Or((\sum_{i=1}^k |\Sys_i| + |\Tys|) \cdot |U|)$
	and results in instances with $n = |U|+k$ vertices, 
	$m = \sum_{i=1}^k |\Sys_i| + |\Tys|$ edges, and parameter $|X| = k$. 
\end{lemma}

\begin{proof}
	Let $(\Hyp,X)$ be the input to \extHS.
	The set $X$ is extendable
	iff there are edges $\{E_x\}_{x \in X} \in \Hyp$ 
	with $E_x \cap X  = \{x\}$ and their union $\bigcup_{x \in X} E_x$ 
	does not contain any edge that is disjoint from $X$.
	This can be phrased as an instance of \mifabbrv by defining,
	for each $x \in X$, the hypergraph $\Sys_x = \{ E \in \Hyp \mid E \cap X = \{x\}\}$.
	The last hypergraph $\Tys$ consists of the edges that are disjoint from $X$.
	Edges that intersect $X$ in more than one vertex can be cast aside.
	This is indeed a linear parameterised reduction.

	For the inverse direction,
	let $(U,\Sys_1, \dots, \Sys_k, \Tys)$ be the instance of \MIF.
	Let $X = \{x_1, \dots, x_k\}$ be a set of $k$ new vertices not previously in $U$.
	We define the hypergraph $\Hyp$ on the vertex set $V = U \cup X$
	by adding all edges of $\Tys$ as well as $E \cup \{x_i\}$
	for every $i \in [k]$ and $E \in \Sys_i$.
	Hypergraph $\Hyp$ can be computed in time 
	\begin{equation*}
		\Or\!\left( \sum_{i=1}^k \sum_{E \in \Sys_i} |E| + \sum_{E' \in \Tys} |E'|\right)
		= \Or\!\left( \left(\sum_{i=1}^k |\Sys_i| + |\Tys| \right) \cdot |U| \right)\!.
	\end{equation*}
	
	For any set $S \subseteq V$, the containment $S \in \Sys_i$ is equivalent to $S \cap X = \{x_i\}$.
	Moreover, the elements of $\Tys$ are exactly those edges of $\Hyp$ that are disjoint from $X$.
	Therefore, there are $S_1 \in \Sys_1, \dots, S_k \in \Sys_k$
	such that $E \nsubseteq \bigcup_{i=1}^k S_i$ holds for all $E \in \Tys$
	if and only if $\{S_i\}_{i \in [k]}$ satisfies \Cref{cond:char_ext_private,cond:char_ext_covering}
	of \Cref{prop:char_extension}, that is, iff $X$ is extendable
	to a minimal hitting set of $\Hyp$.
\end{proof}

The rich structure of \MIF is appreciated when designing algorithms.
For the discussion of its complexity, however,
it is convenient to also have the freedom to choose the sets from a single list.
We thus define the following variant without colours.

\vspace*{.5em}
\noindent
\IF (\ifabbrv)
\begin{description}
	\item [Instance:]Two hypergraph $\Sys$, $\Tys$ on the common vertex set $U$
	
	\hspace*{.9mm}and a non-negative integer $k$.
	\item [Parameter:] The non-negative integer $k$.
	\item [Decision:] Are there $k$ distinct edges $S_1, \dots, S_k \in \Sys$
		such that $\bigcup_{i = 1}^k S_i$
	
		\hspace*{.9mm}does not contain an edge of $\Tys$?
\end{description}
\vspace*{.25em}

\noindent
The two variants are indeed equivalent.

\begin{lemma}
\label{lem:indFam_equivalence}
	\MIF and \IF are equivalent under linear parameterised reductions.	
\end{lemma}

\begin{proof}
	To reduce \mifabbrv to its uncoloured variant,
	it is enough to enforce that selecting two sets of the same colour is never a correct solution.
	They must always cover some forbidden set.
	Let $(\Sys_1, \dots, \Sys_k, \Tys)$ be an instance of \mifabbrv.
	For every index $i\in [k]$, and $S \in \Sys_i$, we introduce a new element $x_{S,i}$.
	The sets are augmented with their respective elements, $S \cup \{x_{S,i}\}$,
	and the results are collected in the single hypergraph $\Sys$.
	Adding the pair $\{x_{S,i}, x_{S',i}\}$ to $\Tys$
	for each $i$ and $S \neq S' \in \Sys_i$ invalidates all unwanted selections.
	It is easy to check that this destroys no valid solution.
	
	For the other direction, we make $k$ copies of $\Sys$
	and ensure that no two copies of the same set are selected together.
	In more detail, we take a new element $x_{S,i}$ for each $S \in \Sys$ and $i \in [k]$,
	define $\Sys_i = \{S \cup \{x_{S,i}\} \mid S \in \Sys\}$, 
	and add the sets $\{x_{S,i}, x_{S,j}\}_{i \neq j}$ to $\Tys\!$.
\end{proof}

In the remainder of this section, we prove that \IF is complete for the class $\W[3]$.
This transfers to \MIF and eventually to \ExtHS via the reductions in \Cref{lem:extension_and_multIndFam,lem:indFam_equivalence}.

\begin{figure}
	\centering
	\includegraphics[scale=.9]{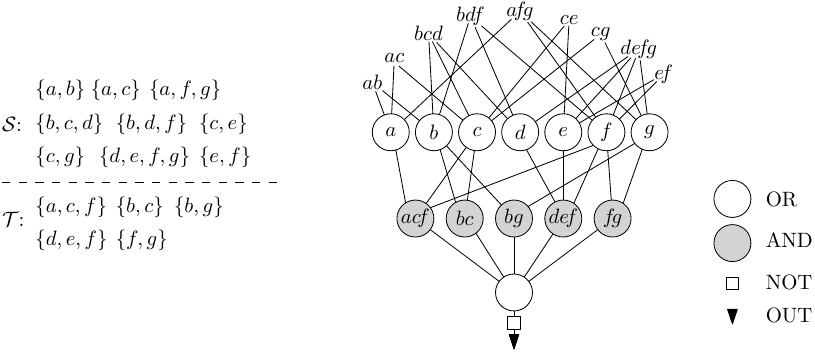}
	\caption{Illustration of \Cref{lem:in_W3}.
		On the left side is an instance of \IF, on the right is the resulting circuit of weft 3.
		All edges are directed downwards.
		Selecting $\{a, c\}$, $\{c,e\}$, and $\{c,g\}$ from $\Sys$ solves the instance for parameter $k=3$,
		any other combination of three sets covers a member of $\Tys\!$.}
	\label{fig:list_clearance_to_circuit}
\end{figure}

\begin{lemma}
\label{lem:in_W3}
	There is a linear parameterised reduction from \IF to the \emph{\textsc{Weighted Circuit Satisfiability}}
	problem on constant-depth circuits of weft $3$.
	In particular, \IF is in $\emph{\W}[3]$.
\end{lemma}

\begin{proof}
	Given an instance $I = (U,\Sys,\Tys,k)$ of \ifabbrv,
	we build a Boolean circuit $C$ of weft $3$
	that has a satisfying assignment of Hamming weight $k$ iff $I$ is a yes-instance.
	\Cref{fig:list_clearance_to_circuit} shows an example instance and the resulting circuit.
	The nodes of $C$ are in one-to-one correspondence to objects in $I$,
	slightly abusing notation we do not distinguish between nodes and their object.
	The input nodes are the edges of $\Sys$. 
	Circuit $C$ has a large OR-gate for each vertex in $u \in U$.
	Node $S \in \Sys$ is wired to gate $u$ whenever $u \in S$.
	Next, we introduce a layer of large AND-gates, one for each forbidden set $E \in \Tys\!$.
	Again, $u$ is connected to $E$ iff $u \in E$.
	The output of all AND-gates lead to a single large OR-gate,
	its \emph{negated} output is the output of $C$.

	Note that the circuit can be constructed from instance $I$ in polynomial time.
	It has depth $4$ and weft~$3$ as every path from an input node to the output
	passes through exactly one large gate in each if the $3$ layers and the (small) NOT-gate.
	We claim that $C$ is satisfied by setting the input nodes $S_1$, \dots, $S_k$ to \true
	if and only if the union $\bigcup_{i=1}^k S_i$ contains no edge of $\Tys\!$.
	
	Let $S_1$ to $S_k$ be a selection of $k$ distinct edges of $\Sys$.
	Assigning \true to the $S_i$ and \false to all others
	satisfies exactly the OR-gates $u \in \bigcup_{i=1}^k S_i$.
	Any AND-gate $E$ of the second layer
	is satisfied iff all its feeding OR-gates are satisfied, 
	that is, iff $E \subseteq \bigcup_{i=1}^k S_i$.
	The results for all forbidden edges $E \in \Tys$ are collected by the large OR-gate in the third layer and subsequently negated.
	Circuit $C$ being satisfied is thus equivalent to
	\textit{no} edge $E$ being contained in the union of $S_1, \dots, S_k$.
\end{proof}

To also show hardness for $\W[3]$, we instead reduce from a problem on Boolean formulas,
that is, circuits in which every gate has fan-out 1.
A formula is called \textit{antimonotone} and \emph{3-normalised}
if it is a conjunction of subformulas in disjunctive normal form (DNF)
with only negative literals.
An example of an antimonotone, 3-normalised formula is
\begin{equation*}
	((\overline x_1 \wedge \overline x_2 \wedge \overline x_4) \vee (\overline x_3 \wedge \overline x_4)) 
		\wedge ((\overline x_1 \wedge \overline x_3) \vee (\overline x_2 \wedge \overline x_5) 
		\vee (\overline x_1 \wedge \overline x_4 \wedge \overline x_5)).
\end{equation*}

\noindent
The example has satisfying assignments of Hamming weight~$0$, $1$, and $2$, but none of larger weight.
The \WANS problem (\wans) is the restriction of \textsc{Weighted Circuit Satisfiability}
to antimonotone, $3$-normalised formulas.
It is complete for the third level of the \mbox{$\W$-hierarchy}~\cite{DowneyFellows13Parameterized,FlumGrohe06ParameterizedComplexityTheory}.

The intuition behind the $\W[3]$-hardness proof is as follows.
The circuit $C$ constructed in \Cref{lem:in_W3} has a single NOT-gate as the output node.
The OR-gates of the first layer are the only ones with fan-out larger than $1$,
but they are connected exclusively to gates of the second layer. 
Moving the negation all the way up to the inputs using De~Morgan's laws,
and duplicating the first layer at most $|\Tys|$ times hence
results in an antimonotone formula that is indeed $3$-normalised.
We show that this is not a mere artefact of the reduction,
but due to a characteristic property of the problem itself.
Namely, every antimonotone, \mbox{$3$-normalised} formula can be
encoded in an instance of the \IF problem.

\begin{figure}
	\centering
	\includegraphics{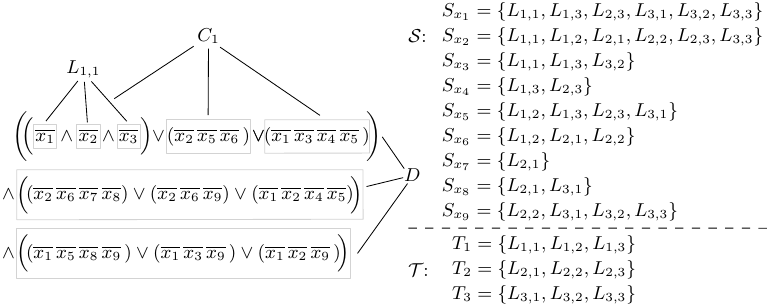}
	\caption{Illustration of \Cref{lem:W3-hardness}.
	On the left side is an antimonotone, 3-normalised formula.
	Negative literals $\neg x_i$ are abbreviated as $\overline{x_i}$
	and conjunctions inside a clause as juxtaposition.
	On the right is the resulting instance of \IF.
	Positions marked with grey boxes are indexed by the respective sets:
	$D$ for the DNF subformulas, $C_1$ for the conjunctive clauses of the first subformula,
	and $L_{1,1}$ for the first clause of the first subformula.
	The formula admits a satisfying assignment of weight $4$ 
	by setting $x_4$, $x_5$, $x_7$, and $x_8$ to \true.
	Equivalently, the union of the sets $S_{x_4}$, $S_{x_5}$, $S_{x_7}$, and $S_{x_8}$ does not cover any forbidden set in $\Tys$.
	No assignment of Hamming weight at least $5$ is satisfying.}
	\label{fig:3_normalised_formula}
\end{figure}

\begin{lemma}
	\label{lem:W3-hardness}
	There is a linear parameterised reduction from \emph{\WANS} to \IF.
	In particular, \IF is hard for $\emph{\W}[3]$.
\end{lemma}

\begin{proof}
	A Boolean formula $\varphi$ on the variable set $\var{\varphi}$ 
	is antimonotone and \mbox{$3$-normalised} if and only if it can be written as
	\begin{equation*}
		\varphi = \bigwedge_{d \in D}\ \bigvee_{c \in C_d}\ \bigwedge_{\ell \in L_{d,c}} \neg {x_{d,c,\ell}},
	\end{equation*}
	
	\noindent
	for an index hierarchy $D$, $\{C_d\}_{d \in D}$, $\{L_{d,c}\}_{d \in D, \nwspace c \in C_d}$,
	and $x_{d,c,\ell} \in \var{\varphi}$.
	The index $d$ ranges over the constituent DNF subformulas,
	$c$ over their conjunctive clauses, and $\ell$ over the negative literals in those clauses.
	Of course, a variable may appear multiple times in the formula,
	so different triples $(d,c,\ell)$ may point to the same variable.
	
	We construct an instance $(U,\Sys,\Tys,k)$ of \MIF
	that is a yes-instance
	if and only if $\varphi$ has a weight-$k$ satisfying assignment.
	This is illustrated in \Cref{fig:3_normalised_formula}.
	We take as vertex set the conjunctive clauses
	$U = \{L_{d,c} \mid d \,{\in}\, D,\, c \,{\in}\, C_d\}$
	and add the edge $S_x = \{L_{d,c} \mid \exists \ell \colon x_{d,c,\ell} \,{=}\, x \}$
	to $\Sys$ for each variable $x \in \var{\varphi}$.
	Namely, $S_x$ contains all clauses in which $x$ occurs.
	The DNF subformulas are represented in the hypergraph $\Tys$
	via the edges $E_d = \{J_{d,c} \mid c \in C_d\}$ for all $d \in D$.
	
	The key observation of this lemma is the following.
	Consider a truth assignment represented by the set $A \subseteq \var{\varphi}$
	of the variables assigned \true.
	Since $\varphi$ is antimonotone, clause $L_{d,c}$ is satisfied if and only if \emph{none} 
	of its variables $x_{d,c,\ell}$ is in $A$.
	As a result, subformula $d$ is \true if and only if $A$
	is \emph{not} a hitting set for the clauses of $d$.
	
	Suppose the assignment $A = \{x_1, \dots, x_k\}$ is satisfying.
	Then, the union $\bigcup_{i=1}^k S_{x_i}$ contains exactly 
	the conjunctive clauses that are not satisfied.
	If this union were to cover any forbidden edge in $\Tys\!$,
	the corresponding subformula, and hence $\varphi$, would be unsatisfied, a contradiction.
	Therefore, $(U,\Sys,\Tys,k)$ is a yes-instance of \IF.
	Conversely, let $S_{x_1}$ through $S_{x_k}$ be a selection of edges from $\Sys$ 
	such that their union covers no member of $\Tys$. 
	In other words, each subformula has at least one clause that is disjoint from $\{x_1, \dots, x_k\}$.
	Assigning \true to (exactly) those variables $k$-satisfies $\varphi$.
\end{proof}

\subsection{Fine-Grained Lower Bounds}
\label{subsec:extension_lower_bounds}

\noindent
We now discuss consequences of our reductions beyond parameterised complexity.
Namely, they allow us to derive lower bounds on the running time of any algorithm
for \ExtHS from certain hypotheses, which are, however, still unproven.

The common believe that the complexity classes \P and \NP are different
can be seen as a conditional (super-polynomial) lower bound on the time complexity of \NP-hard problems.
Similar things can be said about the assumption $\W[1] \neq \FPT$.
Recently, this perspective has been further developed in the area of fine-grained complexity.
It tries to determine the exact exponent of the time needed to solve
various problems in the polynomial, exponential, and parameterised domain.
The proven conditional lower bounds often match closely with the best known algorithmic results,
but they come with the caveat of relying on even more unproven
hardness assumptions.
Such bounds need to strike a balance between the plausibility of the conjecture
and the strength of the result following from it.

We offer three lower bounds on the extension problem.
They are presented in order of increasing strength and
are respectively derived from ever stronger conjectures
about the $\W$-hierarchy and Boolean satisfiability.
The first one immediately follows from \ExtHS being $\W[3]$-complete.
If $\W[3] \neq \FPT$, there is no FPT-algorithm for extension running
in time $f(|X|) \cdot \Or((m{+}n)^{c})$
on hypergraphs with $n$ vertices and $m$ edges
for any computable function $f$ and constant $c$.
Note that the parameterised reductions above also show that \MIF and \IF cannot be solved
respectively in time
$f(k) \cdot \poly(|\Sys_1|, \dots, |\Sys_k|, |\Tys|, |U|)$
and $f(k) \cdot \poly(|\Sys|, |\Tys|, |U|)$.

We derive the second lower bound from the stronger assumption that $\W[2] \neq \FPT$.
For this, we use the following proposition\footnote{%
	The proposition follows from a more general result~\cite[Theorem~4.2]{Chen06LowerBounds}
	on the weighted satisfiability of what the authors call structured $\Pi_t$-circuits.
	For $t=3$, the structure coincides with that of antimonotone, $3$-normalised formulas.
} 
by Chen et al.~\cite{Chen06LowerBounds}.

\begin{proposition}[Chen et al.~\cite{Chen06LowerBounds}]
  \label{prop:WA3NS_and_the_W-hierarchy}
  Let $f$ be an arbitrary computable function.
  If there exists an algorithm solving the
  \emph{\WANS} problem
  on formulas of size $m$ with $n$ variables
  in time $f(k) \, n^{\littleO(k)} \nwspace \emph{\poly}(m)$, then $\emph{\W}[2] = \emph{\FPT}$.
\end{proposition}

\noindent
Note that the reductions from \wans to \ifabbrv,
and further to \mifabbrv and \extHS in
\Cref{lem:W3-hardness,lem:indFam_equivalence,lem:extension_and_multIndFam}
are all polynomial-time computable and
linear the sense that they increase the parameter by at most a constant factor.
In fact, they even preserve the parameter exactly.
Any algorithm solving the \ExtHS problem in time 
$f(|X|) \nwspace (m{+}n)^{\littleO(|X|)}$ on $n$-vertex, $m$-edge hypergraphs
would thus give a fast algorithm for \WANS and thus
imply the collapse $\W[2] = \FPT$.
Similar bounds also hold for the intermediate problems.

The above bound states that the exponent of the worst-case running time for \extHS
necessarily has a linear dependence on the parameter $|X|$.
We show next that the leading coefficient of that dependency is likely to be $1$.
Consider the so-called \textsc{Orthogonal Vectors} (OV) problem as an illustration of this kind of result.
We are given two sets, each with $n$ binary vectors in $d$ dimensions,
and we ought to decide whether there is one vector from each set such that their inner product is $0$.
Straightforwardly testing all pairs yields an $\Or(n^2 \nwspace d)$-time algorithm.
Maybe surprisingly, Williams~\cite{Williams05SETHtoOV} showed that
this cannot be improved to $n^{2-\varepsilon} \cdot \poly(d)$ 
for any constant $\varepsilon > 0$,
at least not if one believes that CNF SAT on formulas with $n$ variables
cannot be solved in time $\Or(2^{(1-\varepsilon/2)n})$.
Such an improved algorithm would be a huge breakthrough in satisfiability,
its conjectured non-existence is the core of the Strong Exponential Time Hypothesis.

We derive our hardness result from a generalisation of OV,
known as $k$-\textsc{Orthogonal Vectors}.
Let $k \ge 2$ be an integer, 
and let $\mathbf{x}_j$ denote the $j$-th component of a vector $\mathbf{x}$.

\vspace*{.5em}
\noindent
$k$-\textsc{Orthogonal Vectors} ($k$-OV)
\begin{description}[labelwidth = 1.8cm]
	\item [Instance:] Sets $A_1, \dots, A_k \subseteq \{0,1\}^d$ with $|A_1| = \dots = |A_k| = n$.
	\item [Decision:] Are there vectors $\mathbf{x}^{(1)} \in A_1, \mathbf{x}^{(2)} \in A_2, \dots, \mathbf{x}^{(k)} \in A_k$
		\vspace*{.4em}
		
		\hspace*{-.5em}such that $\sum_{j = 1}^d  \prod_{i=1}^k \mathbf{x}_j^{(i)} = 0 \nwspace$?
\end{description}

\noindent
The addition and multiplication are those in $\N$, not the field $\mathbb{F}_2$.
We also emphasise that this defines a family of problems, one for each $k \ge 2$,
as opposed to a single parameterised problem.

There are different conjectures
on the hardness of OV and $k$-OV in the literature,
we follow the nomenclature introduced by Gao et al.~\cite{Gao18CompletenessFO}.

\begin{conjecture}[$k$-Orthogonal Vectors conjecture in moderate dimensions]
\label{conj:k-OV}
	For any constants $\varepsilon > 0$ and $k \ge 2$,
	the $k$-\emph{\textsc{Orthogonal Vectors}} problem cannot be solved
	in time $n^{k-\varepsilon} \cdot \emph{\poly}(d)$. 
\end{conjecture}

\noindent
It is well-known that a slight change of the reduction in~\cite{Williams05SETHtoOV}
proves that SETH implies \Cref{conj:k-OV}.
Nevertheless, it is consistent with our current knowledge 
that the \mbox{$k$-OV} conjecture holds while SETH is false.
The assumptions $\W[3] \neq \FPT$ and $\W[2] \neq \FPT$ used above
also follow from SETH but are possibly much weaker, 
see the discussion in~\cite{Chen06LowerBounds,Cygan15ParamertizedAlgorithms,Impagliazzo01ETH1,Impagliazzo01ETH2}.
Again, no inverse connection nor any relation between the conjectures on the \mbox{$\W$-hierarchy}
and on $k$-\textsc{Orthogonal Vectors} are known.

We aim to disprove the existence of an algorithm for \ExtHS running in time
$m^{|X|-\varepsilon} \cdot \poly(n)$ for any constant $\varepsilon > 0$
and constant parameter $|X|$.
By \Cref{lem:extension_and_multIndFam},
such an algorithm implies \MIF being solvable in time
$(\sum_{i=1}^k |\Sys_i| + |\Tys|)^{k-\varepsilon} \cdot \poly(|U|)$.
We show that the latter assertion contradicts the $k$-Orthogonal Vectors conjecture.

\begin{figure}
	\centering
	\includegraphics[scale=.995]{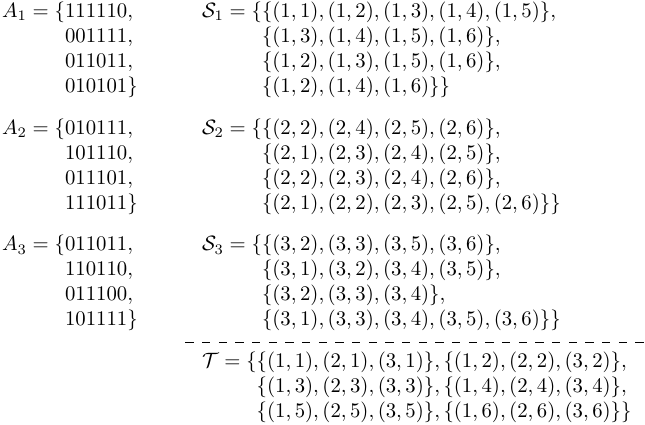}
	\caption{Illustration of \Cref{lem:kOV_to_MIF} for $k = 3$.
	On the left side is an 3-\textsc{Orthogonal Vectors} instance with $n = 4$ vectors in $d = 6$ dimensions.
	On the right is the resulting instance of \MIF.
	The three vectors $010101 \in A_1$, $ 101110 \in A_2$, 
	and $011011 \in A_3$ together are orthogonal, 
	the union of the corresponding edges from $\Sys_1$, $\Sys_2$, and $\Sys_3$
	does not contain any edge of $\Tys$.}
	\label{fig:3OV_to_MIF}
\end{figure}

\begin{lemma}
	\label{lem:kOV_to_MIF}
	If there exists an algorithm solving \MIF in time
	$(\sum_{i=1}^k |\Sys_i| + |\Tys|)^{k-\varepsilon} \cdot \emph{\poly}(|U|)$
	for any constants $\varepsilon > 0$ and $k \ge 2$, 
	then the $k$-OV conjecture fails.
\end{lemma}

\begin{proof}
	Naturally, we reduce from $k$-OV.
	The construction can be seen in \Cref{fig:3OV_to_MIF}.
	Let $A_1, \dots, A_k \subseteq \{0,1\}^d$ be sets with $n$ binary vectors each.
	The constructed instance of \mifabbrv
	has $U = [k] \times [d]$ as its vertex set.
	Let $\mathds{1}(\mathbf{x}) = \{ j \in [d] \mid \mathbf{x}_j = 1 \}$
	be the set with characteristic vector $\mathbf{x}$.
	For each $i \in [k]$, we let the hypergraph $\Sys_i$ represent the set $A_i$
	by adding the edge 
	$\{i\} \times \mathds{1}(\mathbf{x}) = \{(i,j) \mid j \in \mathds{1}(\mathbf{x})\}$
	for each $\mathbf{x} \in A_i$.
	Hypergraph $\Tys$ contains the edge $F_j = [k] \times \{j\}$ for every $j \in [d]$.
	Intuitively, $F_j$ being completely covered by the union of hyperedges
	means that the the corresponding vectors all share a $1$ in the $j$-th component.
	
	Let $\mathbf{x}^{(1)} \in A_1, \dots, \mathbf{x}^{(k)} \in A_k$ be a selection of vectors.
	For any $i \in [k]$ and $j \in [d]$,
	we have $\mathbf{x}^{(i)}_{j} = 0$ if and only if 
	$(i,j)$ is \emph{not} contained in the edge $\{i\} \times \mathds{1}(\mathbf{x}^{(i)}) \in \Sys_i$.
	Moreover, no edge of any other $\Sys_\ell$, $\ell \neq i$, can contain $(i,j)$.
	Therefore, we have $F_j \nsubseteq \bigcup_{i=1}^k (\{i\} \times \mathds{1}(\mathbf{x}^{(i)}))$
	iff $\prod_{i=1}^k \mathbf{x}_j^{(i)} = 0$.
	Finally, this is the case for all $F_j \in \Tys$ iff the vectors
	$\mathbf{x}^{(1)}, \dots, \mathbf{x}^{(k)}$ are orthogonal.
	
	Recall that $k \ge 2$ is a constant.
	There are $\sum_{i=1}^k |\Sys_i| + |\Tys| = kn + d$ edges on $|U| = kd$ vertices
	and the output instance can be computed in time $\Or(knd + kd) = \Or(nd)$.
	Therefore, the assumed algorithm for \MIF running in time 
	$(\sum_{i=1}^k |\Sys_i| + |\Tys|)^{k-\varepsilon} \cdot \poly(|U|)$ 
	would solve the $k$-\textsc{Orthogonal Vectors} instance in time
	\begin{equation*}
		\Or((n+d)^{k-\varepsilon}) \cdot \poly(d)
			 = \Or(n^{k-\varepsilon} + d^{k-\varepsilon}) \cdot  \poly(d)
			 = n^{k-\varepsilon} \cdot \poly(d). \qedhere
	\end{equation*}
\end{proof}

\subsection{The Nondeterministic Strong Exponential Time Hypothesis}
\label{subsec:NSETH}

\noindent
Any algorithm solving the \ExtHS problem in time $m^{|X|-\varepsilon} \cdot \poly(n)$
for arbitrary constants $|X|$ and $\varepsilon >0$ violates SETH.
In \Cref{sec:solving_extension}, we present an $\Or(m^{|X|+1} \nwspace n)$-time solution.
This raises the question what is the ``true'' exponent of $m$.
Although we believe that our algorithm is optimal with respect to $m$,
at least up to subpolynomial factors,
we sketch an argument why it might be hard
to raise the lower bound of \Cref{lem:kOV_to_MIF} to,
say, $m^{|X|+1-\littleO(1)} \cdot \poly(n)$ under SETH.

Carmosino et al.~\cite{Carmosino16NSETH} identified a fundamental obstacle
for proving SETH-hardness.
A co-nondeterministic algorithm for some decision problem 
is one whose computation path may have nondeterministic transitions.
On a yes-instance, every path is required to produce the answer \true,
on a no-instance, there must be at least one path resulting in \false.
The only known co-nondeterministic algorithm for CNF SAT
that improves over brute force is randomized~\cite{Williams16StrongETHBreaksWithMA}.
The \emph{Nondeterministic Strong Exponential Time Hypothesis} (NSETH) conjectures
that this behaviour is inherent to the problem
in that no non-randomized \mbox{co-nondeterministic} algorithm
can break the $2^n$-barrier on formulas with $n$-variables.

\begin{conjecture}[Nondeterministic Strong Exponential Time Hypothesis~\cite{Carmosino16NSETH}]
\label{conj:NSETH}
	For every constant $\varepsilon > 0$, there exists a positive integer $k$
	such that no co-nondeterministic algorithm  without access to randomness 
	can decide $k$\emph{-CNF SAT} on $n$-variable formulas in time $\Or(2^{(1-\varepsilon)n})$.
\end{conjecture}

NSETH can be seen as a common generalisation of SETH and \mbox{$\NP \neq \textsf{co}\NP$}.
The value of the conjecture lies not so much in its plausibility\emDash{}it 
is false for randomized algorithms\emDash{}but the fact that both proving 
and refuting NSETH has interesting consequences.
Finding a fast co-nondeterministic algorithm for satisfiability would immediately
yield new circuit lower bounds, see~\cite{Carmosino16NSETH}.
Proving NSETH would, among other things, resolve the \P vs.~\NP problem.

The conjecture also rules out the existence of certain fine-grained reductions.
Consider a decision problem $\Pi$ that admits an algorithm $\mathcal{A}$ running in time $T(m,n)$
and also a non-randomized co-nondeterministic algorithm $\mathcal{B}$ running in time $T(m,n)^{1-\varepsilon}$ for some constant $\varepsilon > 0$.
If NSETH is true, then no deterministic reduction from CNF SAT to $\Pi$ can prove
that algorithm $\mathcal{A}$ is optimal under SETH
since the very same reduction would give an improved co-nondeterministic algorithm for CNF SAT using algorithm $\mathcal{B}$.

For the further discussion regarding the hardness of \ExtHS,
we use the language of first-order model checking.
For an introduction to first-order logic in parameterised complexity,
see the textbook  Flum and Grohe~\cite{FlumGrohe06ParameterizedComplexityTheory}.
The equivalent \MIF problem can be seen as
deciding whether the input $(U,\Sys_1,\dots,\Sys_k, \Tys)$ is a model\footnote{%
	Strictly speaking, we express the instance $(U,\Sys_1,\dots,\Sys_k, \Tys)$
	as a relational structure over the universe $U \cup \Sys_1 \cup \dots \cup \Sys_k  \cup \Tys$
	with unary relations $S_1, \dots, S_k, T$, where $S_i \nwspace x$ 
	is interpreted as $x$ being an edge of $\Sys_i$,
	and one binary relation 
	$\in \ \subseteq U \times \left(\Sys_1 \cup \dots \cup \Sys_k  \cup \Tys \right)$.
} 
for the formula
\begin{equation*}
	\varphi\ = \ \exists x_1 \in \Sys_1 \dots \exists x_k \in \Sys_k \ \forall y \in \Tys  \ \exists z \in U \colon
		z \in y \wedge \bigwedge_{i = 1}^k z \notin x_i.
\end{equation*}

\noindent
\mifabbrv is a \emph{graph problem} in the sense that the
maximum arity of any relation in $(U,\Sys_1,\dots,\Sys_k, \Tys)$ is $2$.
Formula $\varphi$ has $k$ existential quantifiers, 
followed by a universal one, and then another existential quantifier.
We abbreviate this to $\exists^k \forall \exists$.
Since \mifabbrv is $\W[3]$-complete,
the quantifier structure is a characteristic property of the problem, see~\cite{FlumGrohe06ParameterizedComplexityTheory}.

Let $k$ be a positive integer.
For a graph problem, let $\ell$ denote the total number of ``edges'',
meaning the tuples in the binary relations.
Note that for \mifabbrv $\ell$ can be as large as $(\sum_{i=1}^k |\Sys_i| + |\Tys|) \cdot |U|$.
Along the lines sketched above, Carmosino et al.~\cite[Theorem~4]{Carmosino16NSETH} showed
that under NSETH the \emph{only} graph problems with $k\,{+}\,2$ quantifiers
that can be proven to be SETH-hard with a time bound $\ell^{k+1-\littleO(1)}$ via deterministic reductions are those with quantifier structure $\exists^{k+1}\forall$ or $\forall^{k+1}\exists$.

Using SETH to disprove the existence of an algorithm for \MIF running in time
$\Or(\ell^{k+1-\varepsilon})$, that is, 
\begin{equation*}
	\Or\!\left(\!\left(\!\left(\sum_{i=1}^k |\Sys_i| + |\Tys| \right) |U|\right)^{k+1-\varepsilon}\right)
		= \left(\sum_{i=1}^k |\Sys_i| + |\Tys|\right)^{k+1-\varepsilon} \cdot \poly(|U|),
\end{equation*}
for any $\varepsilon > 0$, would therefore
need to introduce randomness in a non-trivial way
or provide a breakthrough co-nondeterministic algorithm for CNF SAT.

\section{An Algorithm for the Extension Problem}
\label{sec:solving_extension}

\noindent
To finish the description of our hitting set enumeration algorithm,
we need to implement the subroutine for the extension problem.
We not only assumed 
that we can decide for disjoint sets $X$ and $Y$
whether $X$ can be extended to a minimal hitting set avoiding $Y\!$,
we additionally claimed that it is possible to find out whether $X$ is itself a solution
at no additional cost.

\begin{algorithm2e}[t]
\setstretch{1.1}
\vspace*{.25em}
	\Input {Hypergraph $(V,\Hyp)$, $\Hyp \neq \emptyset$, and\\
		\hspace*{1.33cm} disjoint sets $X = \{x_1, \dots, x_{|X|}\}, Y \subseteq V$.}
	\Output{\textsc{minimal} if $X \in \Tr(\Hyp)$,
		\true if there is a $T \in \Tr(\Hyp)$\\ \hspace{1.85cm}with $X \subsetneq T \subseteq V {\setminus} Y\!$, 
		and \false otherwise.}
	\vspace*{.5em}

	\If {$X = \emptyset$}	
	{\label{line:check_non-empty}
		\lIf {$V {\setminus} Y$ is a hitting set} {\Return \true}
		\lElse {\Return \false}
	}
	initialise set system $\Tys = \emptyset$\;	\label{line:initialise_T}
	\lForEach {$x \in X$} {initialise set system $\Sys_x = \emptyset$}
	\ForEach {$E \in \Hyp$}
	{
			\lIf {$E \cap X = \{x \}$} {add $E{\setminus}Y$ to $\Sys_x$}	\label{line:check_for_witnesses}
			\lIf {$E \cap X = \emptyset$} {add $E{\setminus}Y$ to $\Tys$}	\label{line:finish_initialisation}
	}
	\lIf {$\exists \nwspace x \in X \colon \Sys_x = \emptyset$} {\Return \false}	\label{line:everyone_has_a_witness}
	\lIf {$\Tys = \emptyset$} {\Return \textsc{minimal}}	\label{line:end_of_preprocessing}
	\ForEach {$(E_{x_1}, \ldots, E_{x_{|X|}}) \in \Sys_{x_1} \times \dots \times \Sys_{x_{|X|}}$	\label{line:brute_force}}
	{
		$W \gets \bigcup_{i = 1}^{|X|} E_{x_i}$\;
		\lIf{$\forall \nwspace T \in \Tys \colon T \nsubseteq W$} {\Return \true}	\label{line:combine_witnesses}
	} 	
	\Return \false\;	\label{line:end}
\caption{Algorithm for \ExtHS.}
\label[alg]{alg:ExtHS}
\end{algorithm2e}

Despite the hardness results, the investigation in \Cref{sec:extension}
also revealed some structure of the \ExtHS problem that can be exploited algorithmically.
Justified by \Cref{lem:extension_and_multIndFam}, we approach it via \MIF.
Let $\Hyp$ be the input hypergraph.
If $\Hyp = \emptyset$ does not contain a single edge,
$X$ is a minimal transversal if and only if $X = \emptyset$ is empty as well.
In the remainder we assume that $\Hyp$ is non-empty and solve the extension problem 
with \Cref{alg:ExtHS}.
To handle the set $Y$ of excluded vertices,
the algorithm computes the truncated hypergraph $\{E {\setminus} Y \}_{E \in \Hyp}$
and then reduces it to an instance of \mifabbrv.
In fact, both steps can be computed in one pass
(lines~\ref{line:initialise_T}--\ref{line:finish_initialisation}).
\Cref{lem:kOV_to_MIF} suggests
that we cannot improve much over brute force when solving the resulting instance,
at least not in the worst case.
There are, however, several sanity checks possible
that may avoid unnecessary computations in practice.
The first check is the special case of an empty set $X = \emptyset$.
It is extendable without using $Y$ if and only if $V{\setminus}Y$ is a hitting set,
that is, iff $Y$ does not contain an edge.
The other two checks (in lines~\ref{line:everyone_has_a_witness} \& \ref{line:end_of_preprocessing}) assess whether the instance at hand can be decided immediately.
If the checks are inconclusive, 
the instance is indeed solved by brute force (lines~\ref{line:brute_force}--\ref{line:end}).
Note that the existence of a minimal extension is decided without explicitly computing one.
As shown in \Cref{sec:algorithm}, this is enough for the enumeration.

Recall that $n = |V|$ denotes the number of vertices and $m = |\Hyp|$ the number of edges of the hypergraph.
We now show that the running time of \Cref{alg:ExtHS} 
matches the OV-lower bound of \Cref{lem:kOV_to_MIF} up to an $\Or(m)$-factor.

\begin{lemma}
\label{lem:algorithm_ExtHS}
	Let $(V,\Hyp)$ be a non-empty hypergraph
	and $X,Y \subseteq V$ disjoint sets of vertices.
	\Cref{alg:ExtHS} returns \emph{\textsc{minimal}} if $X \in \Tr(\Hyp)$ is a minimal hitting set,
	\emph{\true} if there is a $T \in \Tr(\Hyp)$ with $X \subsetneq T \subseteq V {\setminus} Y\!$, 
	and \emph{\false} otherwise.
	The algorithms runs in $\Or( (\tfrac{m}{|X|})^{|X|} \cdot mn)\!$	time and $\Or(mn)$ space.
\end{lemma}

\begin{proof}
	The first part up to line~\ref{line:end_of_preprocessing} of the algorithm
	computes the reduction from \extHS to \mifabbrv (\Cref{lem:extension_and_multIndFam})
	for the truncated hypergraph $(V {\setminus} Y, \{E {\setminus} Y \}_{E \in \Hyp})$.
	The sanity checks in lines~\ref{line:check_non-empty}, \ref{line:everyone_has_a_witness}, and\ref{line:end_of_preprocessing} filter out trivial instances.
	The foreach-loop starting in line~\ref{line:brute_force} is then brute-forcing
	the result of the reduction, checking all tuples in the Cartesian product $\prod_{x \in X} \Sys_x$.
	
	Since $\Hyp$ is non-empty, 
	the empty set $X = \emptyset$ cannot be a hitting set of $\Hyp$.
	For some $X \neq \emptyset$ to be a hitting set,
	the corresponding hypergraph $\Tys$ must be empty,
	as verified in line~\ref{line:end_of_preprocessing}.
	Observe that this reduces \Cref{prop:char_extension} to \Cref{prop:witnesses}.
	Therefore, such an $X$ is minimal iff every $x \in X$ has a private edge,
	which is exactly what is tested in line~\ref{line:everyone_has_a_witness}.
	In other words, \Cref{alg:ExtHS} correctly identifies the minimal transversals $X$
	and reports this by returning 
	the value \textsc{Minimal} from line~\ref{line:end_of_preprocessing}.
	
	Regarding the time complexity, we assume that all set operations 
	(membership, product, union, intersection, and difference)
	are implemented such that they take time proportional to the total number 
	of elements contained in the input and output of the operation.
	Checking whether $V {\setminus} Y$ is a hitting set and computing the systems $\Sys_{x_{1}}$, \dots, $\Sys_{x_{|X|}}$,
	and $\Tys$ can thus be done in time $\Or(mn)$.
	The running time is dominated by the brute-force phase.
	The cardinality of the Cartesian product is maximum if all systems have the same number of sets
	and no edge is cast aside.
	There are thus at most $(m/|X|)^{|X|}$ many tuples.
	For each of them, the algorithm computes the union $W$ in $\Or(|X| n)$ time 
	and checks all forbidden sets in $\Tys$ in $\Or(mn)$.
	The fact that every element of $X$ has a candidate private edge
	implies $|X| \le m$ and $\Or(|X|n + mn) = \Or(mn)$.

	Regarding the space requirement,
	note that $\Sys_{x}$ and $\Tys$ are all disjoint subhypergraphs of $\{E{\setminus}Y\}_{E \in \Hyp}$,
	using at most as much space as $(V,\Hyp)$.
\end{proof}

Finally, we use \Cref{lem:algorithm_ExtHS} to prove a guarantee on the maximum delay between consecutive
outputs of \Cref{alg:enumerate_recursively}.
The bound is stated in terms of the transversal rank $k^* = \rank(\Tr(\Hyp))$.
Recall that $k^*$ is \textit{not} known to the algorithm,
the input consists only of the hypergraph itself.
For bounded transversal rank, we achieve polynomial delay.
In particular, \Cref{alg:enumerate_recursively} then solves the transversal hypergraph problem in output-polynomial time.

\begin{lemma}
  \label{lem:delay}
  Consider \Cref{alg:enumerate_recursively} with \Cref{alg:ExtHS} implementing the subroutine \Ext.
  On input \mbox{$(V, \order, \Hyp)$}, it enumerates the edges of $\Tr(\Hyp)$ in $\order$-lexicographical order with delay
  $\Or(m^{k^*{+}1} \nwspace n^2)$, where $k^* \,{=}\, \rank(\Tr(\Hyp))$.
  The algorithm uses $\Or(mn)$ space.
\end{lemma}

\begin{proof}
	The correctness was treated in \Cref{lem:enumerate_lexicographically,lem:algorithm_ExtHS}.
	We have also shown there that the label of the current node contains all relevant information
	to govern the tree search.
	In particular, it encodes the path to the node 
	in the (only implicitly constructed) recursion tree for backtracking.
	The total space usage is thus dominated by the $\Or(mn)$ of \Cref{alg:ExtHS}.
	
	We are left to bound the delay.
	The height of the tree is $|V| = n$.
	After exiting a leaf, the pre-order traversal expands at most $2n \,{-}\,1$ inner nodes before arriving at the next leaf.
	In the worst case, method \Ext is invoked in each of them, even with the shortcut evaluations.
	The $\Or((\tfrac{m}{|X|})^{|X|} \nwspace mn) = \Or(m^{|X|+1} \nwspace n)$ subroutine dominates 
	the time spent in each node.
	
	We prove that during the enumeration 
	any set $X$ appearing as the first argument of \Ext is of cardinality at most $|X| \le k^*\!$.
	To reach a contradiction, assume a node $(X,Y,R \,{=}\, V{\setminus}(X {\cup} Y))$ 
	with $|X| \,{>}\, k^*\!$ is expanded by \Cref{alg:enumerate_recursively}.
	This cannot be the root as $X$ is non-empty.
	Thus, prior to entering $(X,Y;R)$, either \Ext{$X\text{,} Y$} has been called
	or the shortcut evaluation inferred the outcome \true from the previous calls.
	Set $X$ is neither a minimal solution nor can it be extended to one
	as its cardinality is larger than the transversal rank.
	The check returned \false and $(X,Y,R)$ is never entered, a contradiction.
	Therefore, the delay is bounded by
	$(2n\,{-}\,1) \,{\cdot} \Or(m^{k^*+1} \nwspace n) = \Or(m^{k^*+1} \nwspace n^2)$.
\end{proof}

\section{Enumerating Unique Column Combinations}
\label{sec:UCC}

\noindent
We apply our enumeration algorithm to hypergraphs arising in data profiling as a proof of concept.
Specifically, we want to solve what is known as the 
the \emph{discovery problem} of minimal unique column combinations.
In data profiling this term is more common than enumeration.
Recall that a UCC for a database $\rel$ over schema $R$ 
is a set $X \subseteq R$ of columns
such that the value combinations appearing as subtuples $r[X]$, $r \in \rel$,
are duplicate-free.

Eiter and Gottlob~\cite{EiterGottlob95RelatedProblems} showed that the minimal
UCCs can be discovered in output-polynomial time
if and only if the transversal hypergraph problem has an output-polynomial solution.
Their proof used a Turing-style reduction that inherently requires exponential space.
Additionally, there is a folklore reduction from the discovery of UCCs to 
the enumeration of the hitting sets of difference sets,
which we sketched in \Cref{subsec:prelims_RelData}.
Intuitively, for any two distinct
rows $r,s \,{\in}\, \rel$, a UCC must contain at least one
attribute in which $r$ and $s$ disagree;
otherwise, the rows are indistinguishable.
That reduction is \emph{parsimonious}\footnote{%
	The concept of parsimonious reductions between enumeration problems is inspired by
	but should not be confused with the homonymous reductions 
	for counting problems~\cite{CapelliStrozecki19Incremental}.
}
in that it establishes a one-to-one correspondence between
the enumeration problems while using only polynomial time and space.
It also preserves set inclusions.
For more details on parsimonious reductions between enumeration problems,
see the work of Capelli and Strozecki~\cite{CapelliStrozecki19Incremental}.
Bläsius, Friedrich, and Schirneck~\cite{Blaesius16DependencyDetection} 
improved upon the Turing-equivalence in~\cite{EiterGottlob95RelatedProblems}
by giving a parsimonious, inclusion-preserving reduction also in the opposite direction,
that is, from hitting sets to UCCs.
Discovering minimal UCCs
is thus exactly as hard as the general transversal hypergraph problem.

This implies a two-phased approach 
for the discovery of UCCs.  
First, generate the hypergraph of minimal difference sets. 
Secondly, list its minimal transversals.
The first phase takes time polynomial in the size of
the database. 
The second phase, which has exponential complexity in the worst case, is the focus of this paper.
In the following, we thus assume that the Sperner hypergraph of minimal difference sets
is given as the input.

\subsection{Data and Experimental Setup}
\label{subsec:UCC_data-exper-setup}

\noindent
We evaluate our enumeration algorithm on a total of 12 databases.
Ten of them are publicly available.
These are the \texttt{abalone}, \texttt{echocardiogram}, \texttt{hepatitis}, and \texttt{horse} datasets
from the University of California Irvine (UCI) Machine Learning Repository;\footnote{\href{https://archive.ics.uci.edu/ml/index.php}{archive.ics.uci.edu/ml}}
\texttt{uniprot} from the Universal Protein
Resource;\footnote{\href{https://www.uniprot.org/}{uniprot.org}}
\texttt{civil\_service},\footnote{\href{https://opendata.cityofnewyork.us}{opendata.cityofnewyork.us}}
\texttt{ncvoter\_allc}\footnote{\href{https://www.ncsbe.gov/index.html}{ncsbe.gov}} and
\texttt{flight\_1k}\footnote{\href{https://transtats.bts.gov/}{transtats.bts.gov}} provided by
the respective authorities of the City of New York, the state of North Carolina,
and the federal government of the United States; \texttt{call\_a\_bike} of the
German railway company Deutsche
Bahn,\footnote{\href{https://data.deutschebahn.com/}{data.deutschebahn.com}} as well as
\texttt{amalgam1} from the Database Lab of the University of
Toronto.\footnote{\href{http://www.cs.toronto.edu/~miller/amalgam/}{dblab.cs.toronto.edu/$\sim$miller/amalgam}}
They are complemented by two randomly generated datasets
\texttt{fd\_reduced\_15} and \texttt{fd\_reduced\_30} using the
\textit{dbtesma} data
generator.\footnote{\href{https://sourceforge.net/projects/dbtesma/}{sourceforge.net/projects/dbtesma}}
Databases with more than ${100}\si{k}$ rows are cut by choosing
${100}\si{k}$ rows uniformly at random.

The algorithms are implemented in C++ and run on a Ubuntu 16.04
machine with two Intel\textsuperscript{\textregistered} Xeon\textsuperscript{\textregistered} E5-2690~v3 \SI{2.60}{GHz} CPUs and \SI{256}{GB} RAM
We made the code and data available.\footnote{
	\href{https://hpi.de/friedrich/research/enumdat.html}{hpi.de/friedrich/research/enumdat}
}
In some experiments, we collect the run times of intermediate steps,
for example the calls to the subroutine (\Cref{alg:ExtHS}).
To avoid interference with the overall run time measurements, we use separate runs for these. 
Also, we average over multiple runs to reduce the noise of the measurements.
See the corresponding sections for details.

\begin{table}[t]
	\begin{center}
	\begin{tabular}{lrrrrrr}
		Dataset & Columns & Rows & $n$ & $m$ & $k^*$ & UCCs \\
		\midrule
		\texttt{call\_a\_bike}   & \SI{17}{}  & \SI{100000}{} & \ \ \SI{13}{} & \SI{6}{}   & \SI{4}{}\ \  & \SI{23}{}	  \\
		\texttt{abalone}		 & \SI{9}{}   & \SI{4177}{}   & \ \ \SI{9}{}  & \SI{30}{}  & \SI{6}{}\ \  & \SI{29}{}     \\
		\texttt{echocardiogram}	 & \SI{13}{}  & \SI{132}{}	  & \ \ \SI{12}{} & \SI{30}{}  & \SI{5}{}\ \  & \SI{72}{}	  \\
		\texttt{civil\_service}	 & \SI{20}{}  & \SI{100000}{} & \ \ \SI{14}{} & \SI{19}{}  & \SI{7}{}\ \  & \SI{81}{}	  \\
		\texttt{horse}			 & \SI{29}{}  & \SI{300}{}	  & \ \ \SI{25}{} & \SI{39}{}  & \SI{11}{}\ \ & \SI{253}{}    \\
		\texttt{uniprot}		 & \SI{40}{}  & \SI{19999}{}  & \ \ \SI{37}{} & \SI{28}{}  & \SI{10}{}\ \ & \SI{310}{}	  \\
		\texttt{hepatitis}		 & \SI{20}{}  & \SI{155}{}	  & \ \ \SI{20}{} & \SI{54}{}  & \SI{9}{}\ \  & \SI{348}{}	  \\
		\texttt{fd\_reduced\_15} & \SI{15}{}  & \SI{100000}{} & \ \ \SI{15}{} & \SI{75}{}  & \SI{3}{}\ \  & \SI{416}{}	  \\
		\texttt{amalgam1}		 & \SI{87}{}  & \SI{50}{}	  & \ \ \SI{87}{} & \SI{70}{}  & \SI{4}{}\ \  & \SI{2737}{}   \\
		\texttt{fd\_reduced\_30} & \SI{30}{}  & \SI{100000}{} & \ \ \SI{30}{} & \SI{224}{} & \SI{3}{}\ \  & \SI{3436}{}   \\
		\texttt{flight\_1k}		 & \SI{109}{} & \SI{1000}{}	  & \ \ \SI{53}{} & \SI{161}{} & \SI{8}{}\ \  & \SI{26652}{}  \\
		\texttt{ncvoter\_allc}	 & \SI{94}{}  & \SI{100000}{} & \ \ \SI{82}{} & \SI{448}{} & \SI{15}{}\ \ & \SI{200907}{} \\
		\\		
	\end{tabular}
	\end{center}
	\caption{
	The databases used in the evaluation, ordered by the number of minimal UCCs.
	Columns and Rows denote the respective dimension of the database, 
	$n$ and $m$ refer to the resulting hypergraph of minimal difference sets,
	$k^*$ is the transversal rank, that is, the size of the largest minimal UCC.}
	\label{tab:data}
\end{table}

\Cref{tab:data} gives an overview of the data.
It lists the number of columns and rows in the database, the number of vertices 
and edges of the resulting hypergraph, 
the transversal rank/maximum cardinality of a minimal UCC,
as well as the number of solutions.
The table is sorted by the
number of minimal hitting sets/UCCs.
All plots below use this order.

After computing the minimal difference sets,
we removed all vertices that do not appear in any edge
as they are irrelevant for the enumeration.
Therefore, the number $n$ of vertices can be smaller than the
number of columns in the database.
The particularly stark difference for \texttt{flight\_1k} stems from a large portions of the columns being empty.
The total number of difference sets of a database with $|r|$ rows
is $\binom{|r|}{2}$ in the worst case.
However, \Cref{tab:data} shows
that the number $m$ of minimal difference sets tends to be much \emph{smaller} than $r$,
let alone quadratic.
Put it the other way around, only very few pairs of rows actually contribute to the UCCs
and the hypergraph perspective thus provides a very compact representation of the discovery problem.
As was observed before by other researchers in data profiling,
the maximum cardinality $k^*$ of the minimal UCCs is small in practice.
In particular, there does not appear to be any relationship between $k^*$ and the input size.

\subsection{Run Time, Delay, and Memory}
\label{subsec:UCC_overall-run-time}

\noindent
Our enumeration method (\Cref{alg:enumerate_recursively})
branches on the vertices in a certain global order.  
Although the order does not matter for our asymptotic bounds, 
it does affect the shape of the explored decision tree,
which in turn impacts the practical run time.
Even on the theoretical side, it has been shown that there exist orders
that render already finding the (lexicographically) first solution
an \NP-hard search problem~\cite{Eiter94ExactTransversal}.

To support the enumeration, 
we heuristically sort the vertices descendingly by the number of distinct values 
that appear in the corresponding column of the original database.
The intuition is that columns with many values have a higher discriminative power over the pairs of rows
and thus are more likely to appear in many minimal UCCs.
Including an expressive vertex makes many other vertices obsolete,
which should lead to early pruning of the tree.
Conversely, excluding such vertices (adding them to the set $Y$ in \Cref{alg:ExtHS})
makes it likely that only a few hitting sets survive, which also prunes the tree early.
Note that reducing the size of the decision tree, and thus the number of subroutine calls,
does not automatically reduce the run time.
The remaining calls may have a larger average return time.
We discuss this in more detail in \Cref{subsec:UCC_oracle-calls}.
As a side note, preliminary experiments showed that sorting the
vertices by their hypergraph degree instead
(that is, the number of minimal difference sets in which they appear)
resulted in similar but slightly worse run times.

\begin{figure}[t]
  \centering
  \includegraphics[width=\textwidth, trim={8 8 8 8}]{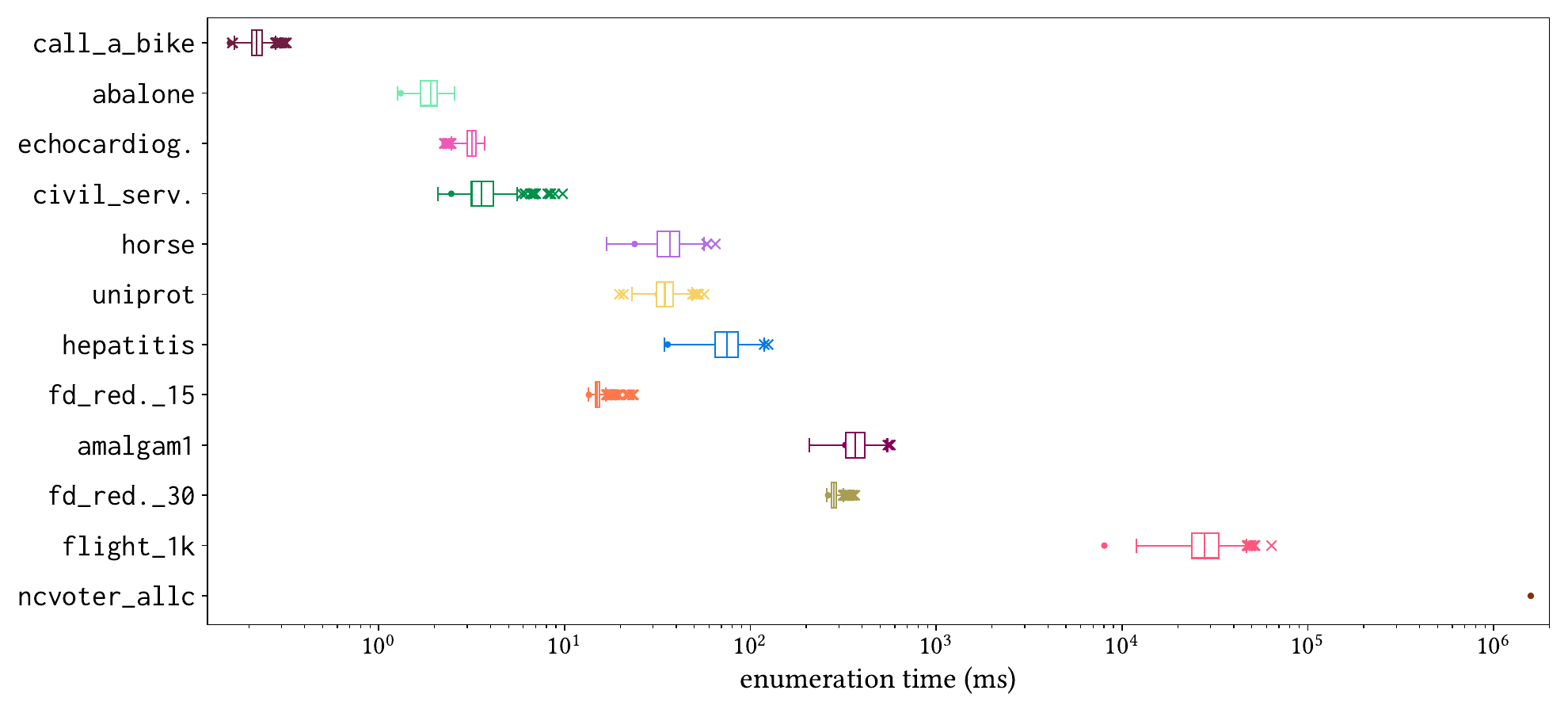}
  \caption{Overall run times of the enumeration algorithm.
	The dot marks the time using the heuristic branching order.
	For all datasets except \texttt{ncvoter\_allc},
	the box plot shows the times for 1000 random orders.
    Their median is indicated by a vertical line,
    the boxes range from the first to third quartile,
    and the whiskers chart the 1.5 interquartile range above and below those quartiles.
    Outliers outside of this range are marked by crosses.
    Each data point is the average over 10 runs.
    }
  \label{fig:run-time}
\end{figure}

Besides using our heuristic order, we also evaluate the algorithms on
1000 random branching orders per dataset.
The \texttt{ncvoter\textunderscore{}allc} instance is an exception
as the larger enumeration times do not permit that many orders.  
We report on \texttt{ncvoter\textunderscore{}allc} separately.
The run times, averaged over 10 measurements for each data point, are shown in \Cref{fig:run-time}.
Note that the \mbox{$x$-axis} is scaled logarithmically.
The boxes show the first to third quartile of the samples, 
with the median indicated as a horizontal line.
The whiskers represent the smallest data point within 1.5 interquartile range (IQR) 
of the lower quartile and the highest one within 1.5 IQR of the upper quartile.
We count everything beyond that as outliers.

The median run times generally scale with the number of solutions, which is to be expected.
They range from $\SI{0.25}{ms}$ for the $23$ minimal UCCs of \texttt{call\_a\_bike}
to roughly $\SI{27}{min}$ for the more than ${200}\si{k}$
solutions of \texttt{ncvoter\textunderscore{}allc}.
The only exceptions from this trend, that have shorter enumeration times albeit more solutions,
are the artificially generated instances
\texttt{fd\textunderscore{}reduced\textunderscore{}15} and
\texttt{fd\textunderscore{}reduced\textunderscore{}30}.
For most of the instances, the branching order had only little impact
and the enumeration times are concentrated around the median.
Our heuristic outperformed the median random order on all instances,
indicating that it is a solid choice in practice. 
On the \texttt{flight\_1k} dataset, 
the heuristic even resulted in a better run time than any of the random orders.
For \texttt{ncvoter\textunderscore{}allc}, however,
the influence of the branching order was significantly larger.
Using the heuristic, the enumeration completed in less than half an hour.
For comparison, on four out of the eleven random orders we tested,
the process only finished after $\SI{59.7}{h}$,
$\SI{105.3}{h}$, $\SI{113.7}{h}$, and $\SI{167.7}{h}$, respectively.
The other seven runs exceeded the time limit of $\SI{168}{h}$
(one week).

\Cref{lem:delay} gives a worst-case guarantee on the delay
that depends on the maximum size $k^*$ of a minimal UCC.
The box plot in \Cref{fig:delays} shows the empirical delays when using the heuristic branching order.
Again, the time-axis is logarithmic.
Recall that the output order of the solutions is entirely determined by the branching order of the vertices.
Each data point in \Cref{fig:delays} corresponds to one output,
it was obtained by averaging the delay prior to the same solution over 100 runs.
The plot shows that there is a high variance in the delays for the different solutions of an instance. 
The extreme case is \texttt{ncvoter\textunderscore{}allc}
where the delays range from the order of $\SI{e-1}{ms}$ to over $\SI{e3}{ms}$.
Nevertheless, the maximum delay was always less than $\SI{2}{s}$, which is reasonably low.
The \texttt{ncvoter\textunderscore{}allc} instance
also has the largest solutions with $k^* = 15$.
However, the next smaller datasets in that category,
\texttt{horse} and \texttt{uniprot} with transversal ranks of $11$ and $10$, respectively,
have a much lower delay.
In general, we cannot confirm a significant correlation between $k^*$ 
and the empirical delays. 
In the following section, we investigate the delays more closely
by looking at the run time distribution of the calls to the subroutine.

\begin{figure}
  \centering
  \includegraphics[width=\textwidth, trim={8 8 8 8}]{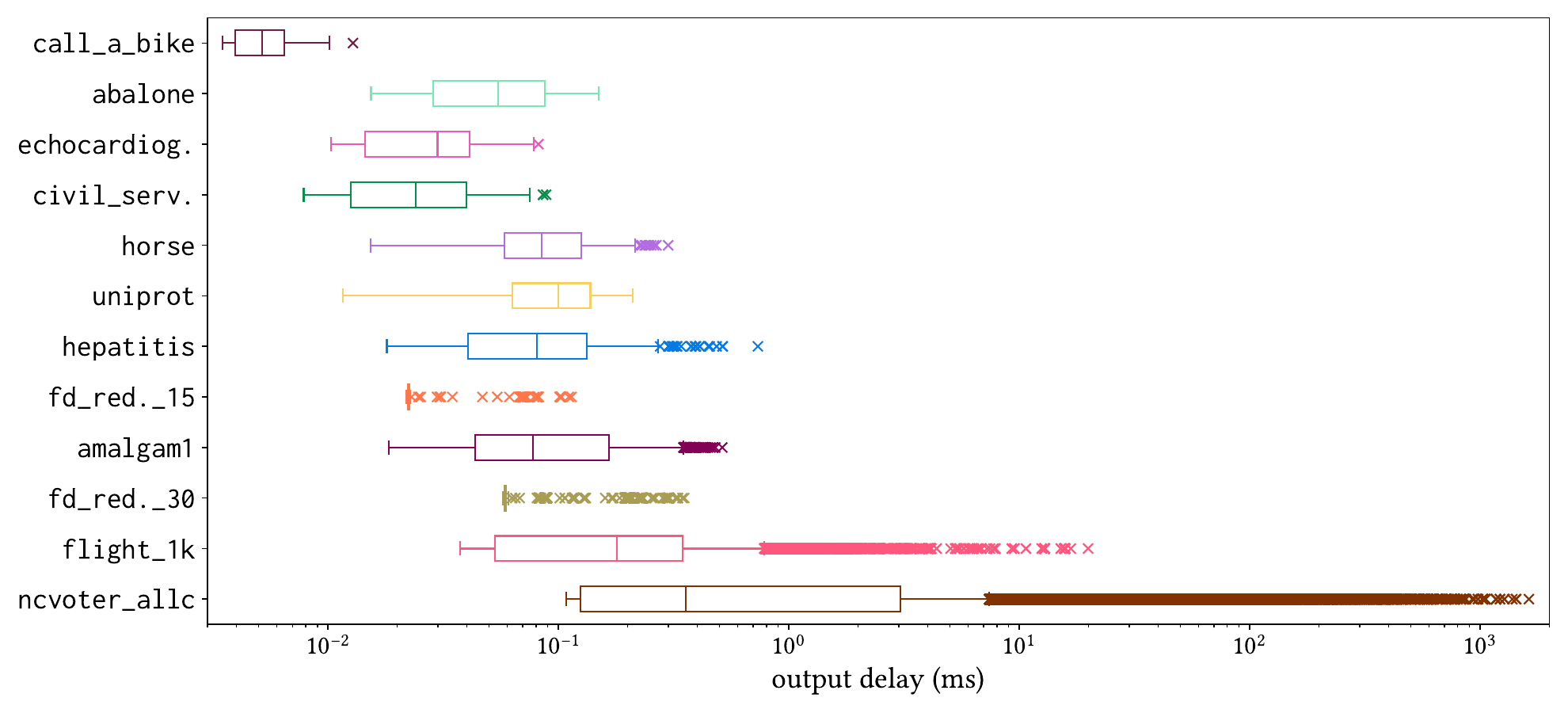}
  \caption{Delays between consecutive outputs of minimal unique column combinations
    using the heuristic branching order.
    The box plots show the same quartiles as in \Cref{fig:run-time}.
    Each data point is the average over 100 runs.
    }
  \label{fig:delays}
\end{figure}

\begin{figure}
    \centering
    \begin{subfigure}{\textwidth}
        \centering
        \includegraphics[width=\textwidth, trim={8 8 8 8}]{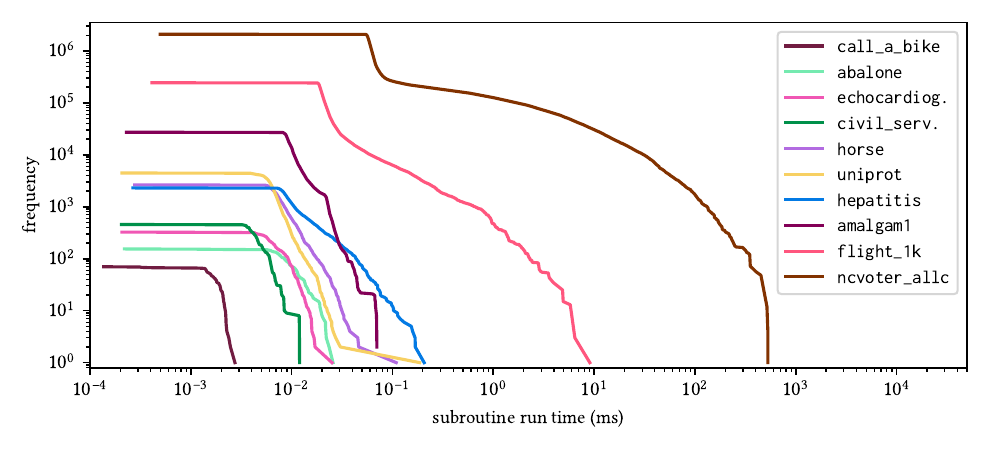}
		\caption{}													
	\label{subfig:all-oracles}
    \end{subfigure}%
	\\
    \begin{subfigure}{\textwidth}
	    \includegraphics[width=\textwidth, trim={8 8 8 8}]{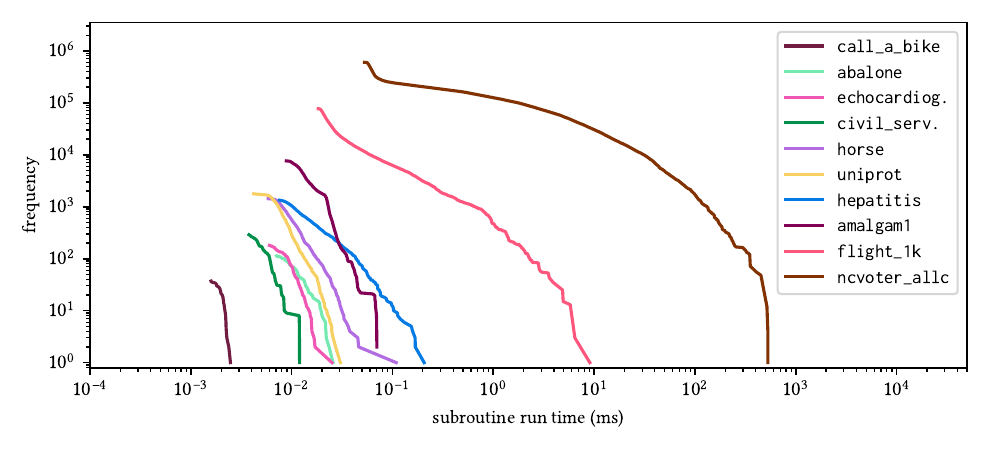}
		\caption{}													
	\label{subfig:oracles-only-brute-force}
    \end{subfigure}
    \caption{The complementary cumulative frequencies
    of the run times of the subroutine calls on the real-world databases
    using the heuristic branching order.
    Plot \subref{subfig:all-oracles} shows all calls, \subref{subfig:oracles-only-brute-force}
  	only those entering the brute-force loop in line~\ref{line:brute_force} of
    \Cref{alg:ExtHS}.
    Each data point is the average of the same call over 100 runs.}
\label{fig:oracle_run_times}
\end{figure}

In \Cref{lem:delay}, we also prove a bound on the space requirement,
which is independent of the number of solutions.
We measured the memory consumption during the enumeration as an average over 5 runs, 
except for \texttt{nc\_voter\_allc} where we did only a single run.
All datasets used between $\SI{4.52}{MB}$
and $\SI{4.68}{MB}$ RAM.
For comparison, just loading the program without an input takes $\SI{4.41}{MB}$.
The memory overhead is marginal and independent of the number of solution.
In our experiments, it even seemed to be insensitive to the given instance.

\subsection{Subroutine Calls}
\label{subsec:UCC_oracle-calls}

\noindent
The only potential reasons for super-polynomial delays 
are the calls to \Cref{alg:ExtHS}.
It is interesting to examine how many calls we need during the enumeration
and how long they actually take in practice.
For our heuristic branching order, we measured the run times of each
individual call, averaged over 100 runs to reduce the noise.
\Cref{fig:oracle_run_times} shows the
\emph{complementary cumulative frequencies} (CCF) of the run times in a log-log plot.
That means, for each time $t$ on the $x$-axis,
the plot shows on the $y$-axis the number of calls with run time at least $t$.
We exclude the artificial instances
\texttt{fd\textunderscore{}reduced\textunderscore{}15} and
\texttt{fd\textunderscore{}reduced\textunderscore{}30} for now,
they are reported separately.
     
First, we examine the impact of the total number of calls on the run time.
The legend of \Cref{fig:oracle_run_times} is ordered by increasing number of solutions,
the same as in the previous plots.
For the real-world databases, this is also the same
as ordering them by increasing \textit{enumeration time}.
For comparison, the total \textit{number} of subroutine calls
is marked by the $y$-value of the left-most endpoint of each curve.
The two orders are almost the same.
An interesting exception is the \texttt{hepatitis} dataset.
It has fewer calls than \texttt{horse} and \texttt{uniprot},
but these calls take more time on average, leading to a higher overall run time.
Instance \texttt{amalgam1} needs even more calls,
which then outweighs the smaller average.
Similarly, the calls for \texttt{horse} take more time than those for
\texttt{uniprot}, but the higher number in the latter case causes a longer run time.
In preliminary experiments, we observed these
effects also when comparing different branching orders for the same dataset.
Aiming for a small number of calls is a good strategy, 
although there are cases where a higher number of easier
calls gives a better result.

Next, we discuss the distribution of the calls. 
The prominent (almost) horizontal lines on the left of \Cref{subfig:all-oracles}
stem from the few trivial calls with $X = \emptyset$.
Those are one to two orders of magnitude faster than 
all other calls since they do not need to construct the instance of \IF.
For the non-trivial cases with $X \neq \emptyset$,
the extension algorithm first checks
whether the resulting instance can already be decided by the sanity checks
in lines~\ref{line:everyone_has_a_witness} and \ref{line:end_of_preprocessing} of \Cref{alg:ExtHS}. 
This way, a significant portion of them can be solved in linear time.
These calls can be seen in the CCF plots as the steep dip immediately following the horizontal lines.
Observe that the $y$-axis is logarithmic,
so the proportion of trivial and easy subroutine calls is actually significant.
Over all databases, slightly more than half of the calls are solved this way.
In fact, for the three instances with the most calls,
namely, \texttt{amalgam1}, \texttt{flight\_1k}, and \texttt{ncvoter\_allc}, 
no more than 32\% of the calls entered the brute-force loop in line~\ref{line:brute_force}.

This loop is the only part of the algorithm that may requires super-polynomial running time.
\Cref{subfig:oracles-only-brute-force} shows the CCFs only for the brute-force calls. 
The differences between \Cref{subfig:all-oracles,subfig:oracles-only-brute-force}
in the lower parts of \texttt{call\_a\_bike} and \texttt{uniprot}
are artefacts of the separate measurements to create these plots.
The run times are heterogeneously distributed with many fast invocations and only a few slow ones. 
As an example, we investigate the calls of the \texttt{flight\_1k} instance.
The database has  \SI{1000}{} rows over \SI{109}{} columns of which 39 are empty
and \SI{17}{} more do not participate in any \emph{minimal} difference set.
The output of \texttt{flight\_1k} are \SI{26652}{} minimal UCCs.
During the enumeration process \Cref{alg:ExtHS} is called \SI{242449}{} times,
\SI{22}{} (0.009\%) calls are trivial, the vast majority of \SI{165767}{} (68.4\%) are
decided easily by the sanity checks, the remaining \SI{76660}{} (31.6\%) calls enter the loop.
Of the brute-force calls, \SI{41353}{} (53.9\%) take only a \emph{single} iteration 
to find a suitable combination of candidate private edges verifying 
that the respective input set $X$ is indeed extendable to
a minimal solution (line~\ref{line:combine_witnesses}).
However, there are also two calls that need the maximum of \SI{74880}{} iterations,
which corresponds to a run time of \SI{16}{ms}.
In those two cases, all possible combinations of potential witnesses had to be tested, only to conclude
that the set is not extendable (line~\ref{line:end}).
It is inherent to the hardness of \ExtHS
that those inputs that are not extendable because all combinations of
candidate private edges cover at least one unhit edge
incur the highest number of iterations and thus longest subroutine run times,
see \Cref{lem:algorithm_ExtHS}.
Fortunately, those occasions were rare in our experiments.
In the case of \texttt{flight\_1k}, only \SI{622}{} calls take more than \SI{10000}{}
iterations, they make up for 0.8\% of the brute-force calls and 0.2\% of all invocations.

The run time distributions for the other real-world databases
are similar to that of \texttt{flight\_1k}, see \Cref{subfig:all-oracles}.
There is always a non-vanishing chance that any given call to the subroutine incurs a high run time,
which is hardly avoidable for a worst-case exponential algorithm,
but even the slowest calls are reasonably fast in practice.
However, the majority of calls is far away from the
worst case, leading to a very low run time on average.
The heterogeneity of the brute-force calls is also showing in the CCFs 
(\Cref{subfig:oracles-only-brute-force}).
They roughly resemble a power-law distribution (straight lines in a log-log plot),
albeit their tendency towards small run times (concavity of the plots)
is stronger than one would expect for a pure power-law.

Another important point of saving related to the subroutine
are those calls that are never actually executed due to the shortcut evaluation
in line~\ref{line:second_Ext} of \Cref{alg:enumerate_recursively}.
We compared the implementation as presented here with a version in which
this optimization is turned off.
Still, the latter version outputs a minimal hitting set as soon as
it is found in line~\ref{line:zeroth_Ext}.
We used the heuristic branching order again.
Over all real-world instances, 
the ratio of calls of the non-optimized version that are skipped by the shortcuts
is between 12.36\% for the \texttt{abalone} dataset
and 66.11\% for \texttt{ncvoter\_allc},
with a median saving of 37.42\%.
The skipped calls are those for which we can be certain that the
given partial solution is indeed extendable, but not yet minimal.
Besides the few calls with $X = \emptyset$,
all of them would have entered the brute-force phase
to find a suitable set of candidate private edges.
On the other hand, they do not need to cycle through all possible
combinations and thus are not the hardest calls.
A given ratio of skipped calls does not directly translate to a certain time saving.
Compared to the enumeration time of the non-optimized version,
the shortcuts gain moderate speedup factors 
from 1.12 for \texttt{abalone} up to 2.26 on the \texttt{uniprot} dataset,
with a median of 1.43.

\begin{figure}[t]
  \centering
  \includegraphics[width=\textwidth, trim={8 8 8 8}]{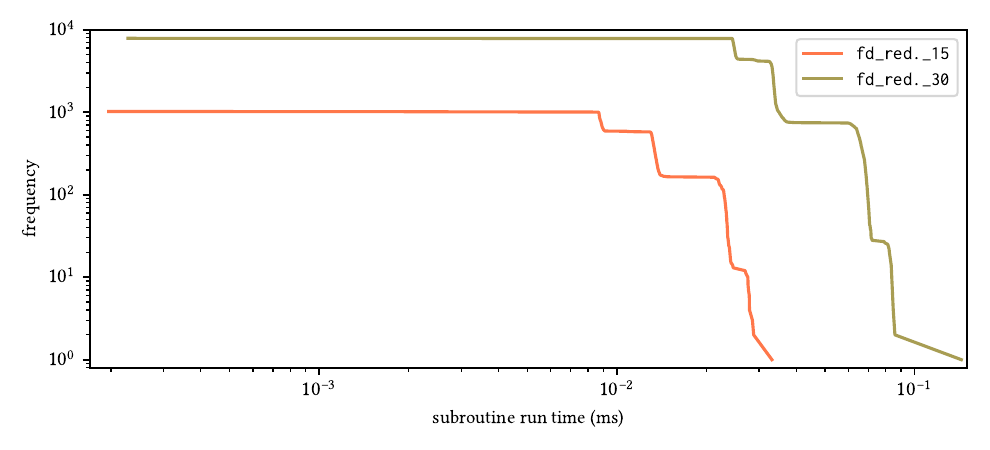}
  \caption{The complementary cumulative frequencies
    of the run times of the subroutine calls on the artificial
    using the heuristic branching order.}
  \label{fig:fd-reduced-oracles}
\end{figure}

Finally, the two artificial instances
\texttt{fd\_reduced\_15} and \texttt{fd\_reduced\_30}
behave very differently from the real-world databases.
\Cref{fig:fd-reduced-oracles} shows their CCFs.
The staircase shape indicates that there are only five types of
calls, with roughly the same run time for all calls of the same type.
Also, the shortcut evaluation hardly saves anything on those datasets.
Only 1.90\% of the calls for \texttt{fd\_reduced\_15} and 
4.97\% for \texttt{fd\_reduced\_30} are skipped,
resulting in a speedup factor of 1.06 on both instances.

\section{Conclusion}
\label{sec:conclusion}

\noindent
We devised a backtracking algorithm for the transversal hypergraph problem
by reducing the enumeration to the \NP-complete decision
whether a set of vertices can be extended to a minimal solution.
Although this may seem counterintuitive, it allowed us to reduce 
both the space usage of the enumeration and the delay.
In particular, we proved that the transversal hypergraph problem 
can be solved simultaneously with polynomial delay and space
on instances whose transversal rank is bounded.
We further showed that the extension problem,
when parameterised by the size of the set to be extended,
is a natural complete problem for the complexity class $\W[3]$.
We presented several conditional lower bounds 
and showed that our extension algorithm is almost optimal under SETH.
With the nondeterministic generalization of SETH, we identified a complexity-theoretic barrier
for closing the remaining gap between our algorithmic results and the lower bound. 

The features of our enumeration method make it particularly suitable for 
the profiling of relational databases,
an application domain where the solutions are expected to be small.
Since the size of the largest solution is the degree of the worst-case time bound,
it could have been that the run times are still prohibitively large in practice.
To guard against such issues, we evaluated our algorithm 
by discovering the minimal unique column combinations 
of several real-world and artificially generated databases.
The experiments showed that our method succeeds within a reasonable time frame,
even when tasked with computing several hundred thousand solutions.

As the empirical run time depends on the branching order of
the vertices, we gave a heuristic that achieves good results 
in practice by reducing the number of calls to the extension subroutine. 
We also verified that the main reason for the low overall run times
is not only the small number of calls but the fact
that the calls are very fast on average.  
In particular,
they regularly avoid the worst case,
which was the basis for the large theoretical bound.

The tree search underpinning our algorithm obviates the need of expensive coordination between branches
or any post-processing of the solutions.
This makes our method easy to implement and memory-efficient.
In particular, approaching the discovery of unique column combinations
as a hitting set problem resulted in an algorithm that does not need
to store previous solutions.
This seems to be a major issue even for current state-of-the-art data profiling algorithms
such as \texttt{DUCC}~\cite{Heise13ScalableUCC} and \texttt{HyUCC}~\cite{Papenbrock17HyUCC}.
Papenbrock and Naumann, the authors of \texttt{HyUCC}, posed the following challenge~\cite{Papenbrock17HyUCC}.

\begin{displayquote}
  For future work, we suggest to find novel techniques to deal with
  the often huge amount of results.  Currently, \texttt{HyUCC} limits
  its results if these exceed main memory capacity [...].
\end{displayquote}

We believe that this can be solved by viewing data profiling from a hitting-set perspective.
However, there are still some problems that need to be overcome to obtain a ready-to-use algorithm.
The enumeration phase is the hard core of the problem,
but it does not seem to be the true bottleneck in practice.
Instead, the quadratic preprocessing step of
preparing the minimal difference sets of the database for our experiments
regularly took much longer than actually enumerating all solutions.
Here, careful engineering has
the potential of huge speedups on real-world instances.
Combining this with the natural advantages of our enumeration algorithm might yield
the novel technique we are looking for.

\vspace*{1em}
\noindent
\textbf{Acknowledgements.}
The authors would like to  thank Felix Naumann and Thorsten Papenbrock 
for the many fruitful discussions about data profiling,
and Erik Kohlros for conducting additional experiments.

\bibliographystyle{plainurl} 
\bibliography{references}

\end{document}